\documentclass{aa}  

\usepackage{graphicx,subfigure}
\usepackage{txfonts}

\usepackage{color}
\usepackage[]{hyperref}
\makeatletter
\renewcommand*\aa@pageof{, page \thepage{} of \pageref*{LastPage}}
\makeatother

\defcitealias{Carretta2009u}{C09u}
\defcitealias{Schiappacasse-Ulloa2024}{SUL24}
\newcommand{\teff}{$T_{\mathrm{eff}}$}
\newcommand{\vm}{V$_{\mathrm{t}}$~}
\newcommand{\msun}{M$_{\odot}$}
\newcommand{\vr}{V$_{\rm r}$~}
\usepackage{longtable}

\begin{document} 

\title{Neutron-Capture Element Signatures in Globular Clusters:}
\subtitle{Insights from the {\it Gaia}-ESO Survey}

   \author{J. Schiappacasse-Ulloa \inst{1}
          \and
          L. Magrini \inst{1}
          \and
          S. Lucatello \inst{2}
          \and
          S. Randich \inst{1}
          \and
          A. Bragaglia \inst{3}
          \and
          E. Carretta \inst{3}
          \and
          G. Cescutti \inst{4,5,6}
          \and
          F. Rizzuti  \inst{5,6,7}
          \and
          C. Worley   \inst{8,9}
          \and
          F. Lucertini \inst{10}
          \and
          L. Berni \inst{1,11}}
    \institute{INAF–Osservatorio Astrofisico di Arcetri, Largo Enrico Fermi 5, 50125 Florence, Italy\\
              \email{jose.schiappacasse@inaf.it}
         \and
             INAF–Osservatorio Astronomico di Padova, Vicolo dell’Osservatorio 5, 35122 Padova, Italy
        \and
            INAF-Osservatorio di Astrofisica e Scienza dello Spazio di Bologna, via P. Gobetti 93/3, 40129, Bologna, Italy
        \and
            Dipartimento di Fisica, Sezione di Astronomia, Università di Trieste, Via G. B. Tiepolo 11, 34143, Trieste, Italy
        \and
            INAF – Osservatorio Astronomico di Trieste, Via G.B. Tiepolo 11, 34143, Trieste, Italy
        \and
            INFN, Sezione di Trieste, Via A. Valerio 2, 34127, Trieste, Italy
        \and
            Heidelberger Institut für Theoretische Studien, Schloss-Wolfsbrunnenweg 35, D-69118 Heidelberg, Germany
        \and
            Institute of Astronomy, University of Cambridge, Madingley Road, Cambridge, CB3 0HA, UK
        \and
            School of Physical and Chemical Sciences – Te Kura Matū, University of Canterbury, Private Bag 4800, Christchurch, 8140, New Zealand
        \and
            ESO – European Southern Observatory, Alonso de Cordova 3107, Vitacura, Santiago, Chile
        \and
            Dipartimento di Fisica e Astronomia, Università degli Studi di Firenze, Via Sansone 1, 50019, Sesto Fiorentino, Italy
             }

   \date{Received NN 2025; accepted NN 2025}

 
  \abstract
   {Globular clusters (GCs) are crucial to our understanding of the formation and evolution of our Galaxy. While their abundances of light and iron-peak elements have been extensively studied, research on heavier elements and  their possible link to both the multiple stellar population phenomenon and the origin of globular clusters remains relatively limited.} 
   {Our aim is to analyse the chemical abundances of various neutron-capture elements, using GCs as tracers of the Galactic halo. Furthermore, we explore the potential connection between these elements and the multiple stellar population phenomenon in GCs to better constrain the nature of the polluters responsible for the intracluster enrichment. Additionally, we seek to distinguish the origins of GCs based on their neutron-capture element abundances.}
   {We analysed a sample of 14 GCs spanning a wide metallicity range,  [Fe/H] from -0.40 to -2.32, observed as a part of the {\it Gaia}-ESO survey and analysed using a homogeneous methodology. Specifically, here we present results for Y, Zr, Ba, La, Ce, Nd, Pr, and Eu, obtained from FLAMES-UVES spectra. We compared our results with a stochastic Galactic chemical evolution model. }
   {Except for Zr, the Galactic chemical evolution model describes well the broad trend displayed by neutron-capture elements in GCs, when it was available. 
   Moreover, in some clusters, a strong correlation between hot H-burning (Na, Al) and s-process elements suggests a shared nucleosynthetic site, e.g., asymptotic -giant branch (AGB) stars of different masses and/or fast-rotating massive stars that produced the intracluster pollution. 
   Additionally, we identified clear differences in the [Eu/Mg] ratio between in-situ ($\langle$[Eu/Mg]]$\rangle$=0.14 dex) and ex-situ ($\langle$[Eu/Mg]]$\rangle$=0.32 dex) GCs, revealing their distinct chemical enrichment histories. Finally, on average, the Type II GCs NGC~362, NGC~1261, and NGC~1851 show a spread ratio in s-process elements between second- and first-generation stars roughly twice as large as that observed in Type I clusters.}
   {}

   \keywords{stars: abundances -- stars: Population II -- (Galaxy:) globular clusters: general -- (Galaxy:) globular clusters: individual: (NGC~104, NGC~362, NGC~1261, NGC~1851, NGC~1904, NGC~2808, NGC~4372, NGC~4590, NGC~4833, NGC~5927, NGC~6218, NGC~6752, NGC~7078, NGC~7089)}

   \maketitle
%

\section{Introduction}
\label{introduction}

Over the past decades, GCs have been the subject of extensive research. Their age and spatial distribution make them perfect candidates for studying the formation and evolution of the Galactic halo, disc, and bulge. 
With some exceptions \citep[e.g., the GC E~3][]{Salinas2015,Monaco2018}, all the known Galactic GCs host multiple stellar populations (MPs), a phenomenon which is generally attributed to internal pollution caused by a subset of stars, indicated as polluters \citep[see e.g.,][]{Bastian2018,Gratton2019}. While the first generation (FG) of stars displays a chemical pattern similar to field stars of the same metallicity, the second generation (SG) shows an altered composition in hot H-burning elements (specifically, a high content of N, Na, and Al, but a depletion in C, O, and Mg). Although the presence of MPs in GCs is well established, it is still unclear what the sources of such altered composition in SG stars are. Different mechanisms have been proposed, such as, e.g.,  intermediate-mass (4-8\msun) AGB stars \citep{Ventura2001}, fast-rotating massive stars \citep{Meynet2006}, massive binaries \citep{DeMink2009}, super-massive stars \citep{Denissenkov_2015}, and non-canonical stellar evolution, including rotation and binarity \citep{Pancino2018A&A...614A..80P}. 
Although MPs have been extensively studied, most spectroscopic analyses focus primarily on light and Fe-peak elements. So far, these observations have not been sufficient to determine the nature of the polluters.
In this regard, a more comprehensive chemical characterisation of other key elements, such as Li \citep[see, e.g.][]{D'orazi2010_Li, D'orazi2014, D'orazi2015, Schiappacasse-Ulloa2022}  and neutron-capture (n-capture) elements \citep[see, e.g.][]{dorazi2010_neutron,Cohen2011, Schiappacasse-Ulloa2023}, could provide deeper insights into the nature of the polluters. 

In this respect, n-capture elements can be broadly classified into two categories based on their formation processes: rapid- (r-) and slow- (s-) process elements. The primary astrophysical sites of r-process nucleosynthesis remain a topic of debate. However, it is generally accepted that short-timescale sources, such as magnetorotationally driven supernovae (MRD SNe; \citealt{Nishimura2015}) or collapsars (\citealt{Siegel2019}), contribute significantly to their production. Additionally, the merger of binary neutron star systems is expected to play a crucial role in r-process enrichment over longer timescales \citep[see, e.g.,][]{Perego2021, Cowan2021RvMP...93a5002C}. On the other hand, the s-process happens in a variety of sources such as low-mass (1.2-4\msun) AGB stars (with some contribution of AGB of up to 8\msun) or rotating massive stars, which are efficiently producing the first-peak s-process elements such as Y and Zr \citep[][]{Frischknecht2012,Limongi2018} and, to some degree, also second-peak s-process elements such as Ba, La, and Ce. Although the importance of these elements is well recognised in the identification of relevant nucleosynthetic sites, only a limited number of studies have been conducted in GCs \citep[see e.g.,][]{James2004}. These studies have revealed a relatively homogeneous internal abundance of n-capture elements in the analysed GCs \citep[see, e.g.][]{dorazi2010_neutron,Cohen2011}. Recently, \citet{Schiappacasse-Ulloa2024} (hereinafter \citetalias{Schiappacasse-Ulloa2024}) conducted a homogeneous analysis of Y, Ba, La, and Eu abundances in 18 globular clusters using UVES spectra, representing the largest and most comprehensive sample analysed to date. Among their most remarkable results, they found evidence of a slight correlation between Y and Na in GCs with intermediate metallicities (-1.8<[Fe/H]<-1.1 dex), suggesting a potential modest production of light s-process elements alongside Na production.

\citet{Milone2017} classified GCs into two types based on their chemical properties. Type I GCs typically show uniform n-capture element abundances across their stellar populations, despite minor iron variations. In contrast, Type II GCs exhibit more complex chemical patterns, often including significant spreads in iron and heavier elements such as those produced by the n-capture process. Similar to those observed in satellite galaxies, these chemical peculiarities could reflect their formation in distinct environments. 
Recently, efforts have been made to explore the potential chemical differences between GCs formed within our Galaxy (in-situ GCs) and those born outside and later accreted (ex-situ GCs). For example, \citet{Carretta2022} suggested that some of the GCs born in the Sagittarius dwarf galaxy could be distinguished by their Fe-peak abundance pattern, which displays lower Sc, V, and Zn than in-situ stars. More recently, \citet{Monty2024} analysed a sample of 54 ex-situ and in-situ GCs, finding a clear difference in their [Eu/Si] ratio. Specifically, accreted GCs exhibit elevated [Eu/Si] ratios, whereas in-situ GCs show lower values.

In this regard, extensive spectroscopic surveys, such as  the {\it Gaia}-ESO survey \citep{Gilmore2022,Randich2022}, the Apache Point Observatory Galactic Evolution Experiment \citep[APOGEE;][]{Majewski2017} and the GALactic Archaeology with HERMES \citep[GALAH;][]{Buder2019} are significantly advancing our understanding of Galactic stellar populations and, in particular, of GCs.
Specifically, {\it Gaia}-ESO is the only large public survey to benefit from an 8-m class telescope and a spectrograph that achieves a resolving power R=47,000 \citep[][FLAMES equipped with UVES]{Pasquini2002Msngr.110....1P}. 
These characteristics, along with the broad wavelength coverage (480.0-680.0~nm), the high signal-to-noise (S/N), and homogeneous spectral analysis, enabled {\it Gaia}-ESO to perform a precise determination of a large number of element abundances simultaneously, including a large number of n-capture elements \citep[see, e.g.][]{Magrini2018A&A...617A.106M, Van2023A&A...670A.129V, Molero2023, Molero2024arXiv241211844M}. Consequently, the {\it Gaia}-ESO data have the potential to enhance our understanding of the internal chemical enrichment of globular clusters while also providing valuable insights into their role in the formation and evolution of the Galaxy.

In the present work, we built upon the analysis conducted by \citetalias{Schiappacasse-Ulloa2024}, expanding both the sample and the species analysed. We provided a homogeneous analysis of a large sample of GCs covering the full range of metallicities. We aimed to examine their internal spreads, explore their relationship with key elements in the context of the MP phenomenon —such as hot H-burning elements— and investigate any potential links to their origin.
We organised the article as follows: In \S\ref{Sec:Sample_Membership}, we described the selection of GC members. In \S\ref{Sec:DataAnalysis}, we detailed the data reduction and abundance determination. The \S\ref{Sec:YBA_Vm} and the Appendix \S\ref{Sec:comp_lit} compared our results with the literature for in-common stars and the Appendix \S\ref{App:vm} describes trends in our results with microturbulent velocity, \vm. In \S\ref{Sec:Result_GalacticContext}, we analysed the {\it Gaia}-ESO results in the Galactic context by using the predictions of an updated stochastic Galactic chemical evolution model. Moreover, we studied the n-capture relation to the MP indicators in \S\ref{Sec:n-cap_vs_NAL} and the Appendix \S\ref{App:ncap-MP}, and in \S\ref{Sec:n-cap_GCs}, the internal spread of n-capture abundances in our sample GCs. In \S\ref{Sec: Summary}, we summarised our results. 

\begin{table*}
\centering
\caption{GCs sample with its respective position, metallicity, \vr was taken from {\it Gaia}-ESO \citep{Gilmore2022,Randich2022}, the astrometric information (proper motions) and parallaxes were taken from {\it Gaia} e{\sc dr3}\citep{GaiaCollaboration2021}. We also list the lowest membership probability (MEM3D; when available) among the selected cluster members. The age column lists the GCs age reported by (1)\citet{Forbes2010} or (2)\citet{DeAngeli2005}. The last column displays the GCs origin reported by \citet{Massari2019} being (3)~in-situ and (4)~ex-situ.}
\resizebox{\textwidth}{!}{
\begin{tabular}{ccccccccccc}
\hline
\hline
Cluster       & RA          & DEC         & {[}Fe/H{]} & PMRA & PMDEC & \vr     & Parallax & MEM3D & Age & Origin\\
              & J2000.0     & J2000.0     &     dex    & mas yr$^{-1}$ &  mas yr$^{-1}$     & km s$^{-1}$  &  mas & & Gyr  & \\
\hline
NGC~104  & 00 24 05.67 & -72 04 52.6 & -0.79$\pm$0.07  &  5.30$\pm$0.01  &   -2.61$\pm$0.02  & -18.7$\pm$0.4 &  0.21$\pm$0.01 & 0.99 & 11.75$^{1}$ &  3\\
NGC~362  & 01 03 14.26 & -70 50 55.6 & -1.13$\pm$0.06  &  6.73$\pm$0.02  &   -2.57$\pm$0.02  & 223.2$\pm$0.3 &  0.08$\pm$0.01 & 0.97 & 10.75$^{1}$ &  4\\
NGC~1261 & 03 12 16.21 & -55 12 58.4 & -1.21$\pm$0.06  &  1.61$\pm$0.02  &   -2.07$\pm$0.02  &  73.9$\pm$0.2 &  0.03$\pm$0.02 & 0.99 & 10.75$^{1}$ &  4\\
NGC~1851 & 05 14 06.76 & -40 02 47.6 & -1.10$\pm$0.09  &  2.16$\pm$0.02  &   -0.68$\pm$0.02  & 321.8$\pm$0.3 &  0.04$\pm$0.01 & NaN  & 11.00$^{1}$ &  4\\
NGC~1904 & 05 24 11.09 & -24 31 29.0 & -1.59$\pm$0.05  &  2.45$\pm$0.02  &   -1.61$\pm$0.02  & 205.3$\pm$0.4 &  0.06$\pm$0.02 & 0.99 & 10.25$^{2}$ &  4\\
NGC~2808 & 09 12 03.10 & -64 51 48.6 & -1.09$\pm$0.07  &  1.01$\pm$0.02  &    0.31$\pm$0.02  & 103.4$\pm$0.4 &  0.09$\pm$0.01 & 0.94 & 11.00$^{1}$ &  4\\
NGC~4372 & 12 25 45.40 & -72 39 32.4 & -2.10$\pm$0.17  & -6.38$\pm$0.01  &    3.19$\pm$0.01  &  79.0$\pm$0.4 &  0.17$\pm$0.01 & NaN  & 12.50$^{2}$ &  3\\
NGC~4590 & 12 39 27.98 & -26 44 38.6 & -2.26$\pm$0.11  & -2.73$\pm$0.02  &    1.71$\pm$0.02  & -92.0$\pm$0.4 &  0.06$\pm$0.02 & NaN  & 12.00$^{1}$ &  4\\
NGC~4833 & 12 59 33.92 & -70 52 35.4 & -1.93$\pm$0.08  & -8.39$\pm$0.01  &   -0.99$\pm$0.02  & 200.5$\pm$0.4 &  0.12$\pm$0.01 & NaN  & 12.50$^{1}$ &  4\\
NGC~5927 & 15 28 00.69 & -50 40 22.9 & -0.40$\pm$0.06  & -5.15$\pm$0.05  &   -3.11$\pm$0.06  &-102.2$\pm$0.4 &  0.06$\pm$0.06 & 0.97 & 10.75$^{1}$ &  3\\
NGC~6218 & 16 47 14.18 & -01 56 54.7 & -1.33$\pm$0.08  & -0.21$\pm$0.02  &   -6.83$\pm$0.01  & -41.2$\pm$0.3 &  0.17$\pm$0.02 & NaN  & 13.00$^{1}$ &  3\\
NGC~6752 & 19 10 52.11 & -59 59 04.4 & -1.55$\pm$0.07  & -3.18$\pm$0.02  &   -3.98$\pm$0.02  & -25.9$\pm$0.4 &  0.23$\pm$0.03 & 0.99 & 12.50$^{1}$ &  3\\
NGC~7078 & 21 29 58.33 & +12 10 01.2 & -2.32$\pm$0.11  & -0.67$\pm$0.02  &   -3.75$\pm$0.01  &-106.6$\pm$0.3 &  0.06$\pm$0.02 & NaN  & 12.75$^{1}$ &  3\\
NGC~7089 & 21 33 27.02 & -00 49 23.7 & -1.42$\pm$0.17  &  3.35$\pm$0.03  &   -2.20$\pm$0.01  &   0.1$\pm$0.4 &  0.04$\pm$0.02 & NaN  & 11.75$^{1}$ &  4\\
\hline
\hline
\end{tabular}}
\label{tab:gcs_info}
\end{table*}

\section{Sample and Membership Selection}
\label{Sec:Sample_Membership}

The {\it Gaia}-ESO survey observed 14 GCs as calibrators of both stellar parameters and chemical abundances. The sample of GCs, together with the other stellar calibrators - open clusters, benchmark stars, asteroseismic targets - is described in \citet{Pancino2017A&A...598A...5P}, while the results of their analysis, in the context of the global calibration and homogenisation of the survey, are reported in \citet{Hourihane2023A&A...676A.129H}. The GCs were selected to cover a wide metallicity range, extending from about [Fe/H]$=-2.5$ to $-0.5$.  Their data have been used in various scientific studies, such as, e.g., \citet{Lardo2015A&A...573A.115L},  \citet{Pancino2017A&A...601A.112P} and \citet{Tautvai2022A&A...658A..80T}.

In our work, we accounted for the complete sample of GCs, using only the high-resolution UVES spectra with signal-to-noise ratio larger than 50\footnote{median signal-to-noise per pixel over the whole spectrum.}.
To build our sample, we selected stars from the final {\it Gaia}-ESO release \citep{Hourihane2023A&A...676A.129H} that were observed with UVES and belonged to the GC fields. The latter condition corresponds to GES\_TYPE = \textit{GE$\_$SD$\_$GC} and \textit{AR$\_$SD$\_$GC}, the former for new {\it Gaia}-ESO observations and the latter for archival data re-analysed in {\it Gaia}-ESO.
This condition does not guarantee that such stars are actually members of GCs. We therefore used the membership information contained in the MEM3D column of the Gaia-ESO catalogue, when available, to determine whether a star was a member of a GC. For the determination of the probability of membership in the {\it Gaia}-ESO survey, based both on {\it Gaia} proper motions and parallaxes and on {\it Gaia}-ESO radial velocities, we refer the reader to \citet{Jackson_2022}. When the membership probability value was available (for seven GCs), we selected stars with MEM3D > 0.9. 

\begin{figure*}
        \centering
        \includegraphics[width=\textwidth]{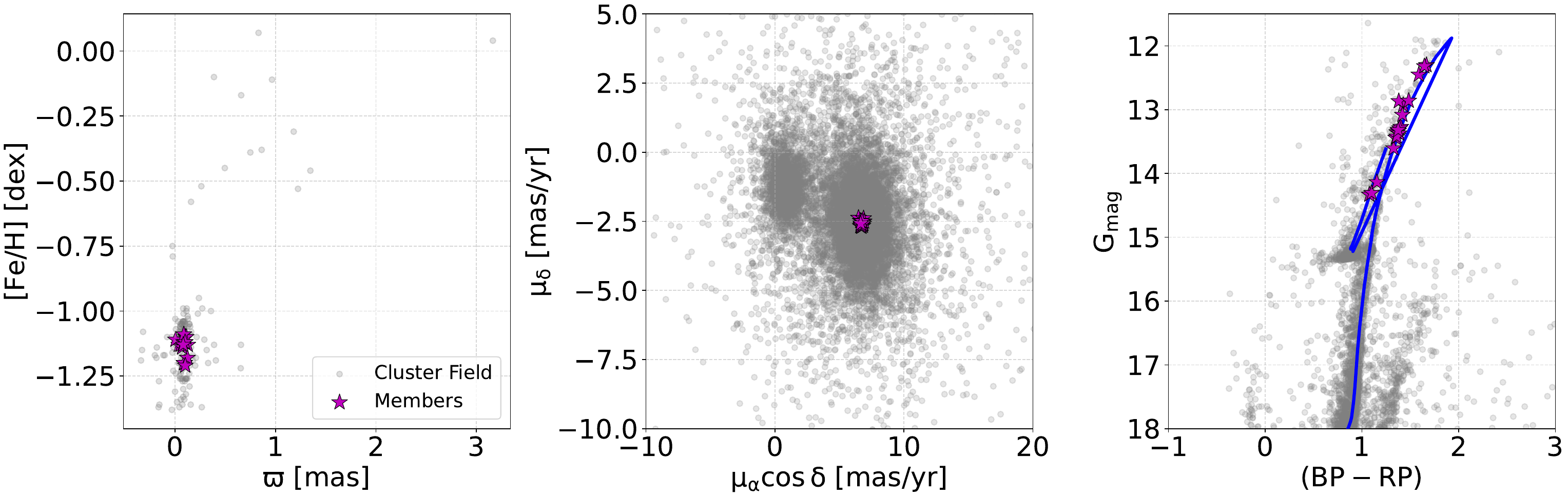}
        \caption{Membership selection for NGC~362. The figure display the properties of members and non-members, represented with magenta and grey dots, respectively. Different panels show different quantities and  selection criteria going from left to right: [Fe/H]-parallax, proper motions, and colour-magnitude diagram. On the latter panel, we overplot in blue the corresponding isochrone corresponding to [Fe/H]=-1.13 dex, [$\alpha$/Fe], E(B-V)=0.05, (m-M)$V$=14.83, and age of 10.75 Gyr.}
        \label{fig:mem_n362}
\end{figure*}

For the remaining seven clusters, we performed a $\sigma$-clipping process to clean our sample from field stars taking into account their parallax ($\overline{\omega}$), proper motions, radial velocity from {\it Gaia} e{\sc DR3} data \cite{GaiaCollaboration2021}, [Fe/H] and position on the colour-magnitude diagram (considering only those consistent with the isochrone for the cluster age and metallicity from literature data). As a sanity check, we compared our membership selection with the catalogue from \citet{Vasiliev2021}, which used {\it Gaia} e{\sc DR3} data to study the kinematic properties of GCs and computed membership probabilities. All the stars we selected had a membership probability of at least 0.99 in that catalogue.
An example of member selection is shown in Figure \ref{fig:mem_n362} in which we display the selected stars in NGC~362 in the planes: [Fe/H] versus parallaxes, proper motions, and location in the colour-magnitude diagram. In addition, we overplotted the PARSEC isochrone \citep{Bressan2012} corresponding to [Fe/H]=-1.13 dex, E(B-V)=0.05 and (m-M)$_V$=14.83 \citep{Harris2010}, age of 10.75 Gyr \citep{Forbes2010}, and assuming an [$\alpha$/Fe]=0.25 dex. Table \ref{tab:gcs_info} displays the information about the coordinates (RA and DEC), [Fe/H] and \vr from {\it Gaia}-ESO \citep{Gilmore2022,Randich2022}, proper motions and parallaxes from {\it Gaia} e{\sc DR3}, and ages (from \cite{Forbes2010} or \cite{DeAngeli2005}) of the final GC sample.

\section{Data analysis and abundances}
\label{Sec:DataAnalysis}

We utilised data from the {\sc dr5.1} release of the {\it Gaia}-ESO Survey\footnote{\hyperlink{https://www.eso.org/qi/catalog/show/411}{https://www.eso.org/qi/catalog/show/411}}, which were obtained through spectral analysis of UVES spectra (with a resolving power of R = 47,000 and a spectral range of 480.0–680.0~nm). The spectra were reduced and analysed by the {\em Gaia}-ESO consortium, with the data processing organised into various nodes. Each node applied different pipelines for spectral analysis, and their results were subsequently combined. To ensure consistency, the outputs from all WGs were homogenised using a calibrator database, which includes benchmark stars and open or globular clusters. This calibration strategy was based on the approach outlined by \citet{Pancino2017A&A...598A...5P} and was further refined by WG15 \citep{Hourihane2023A&A...676A.129H}.

The team at INAF–Osservatorio Astrofisico di Arcetri carried out the data reduction, as well as the determination of radial and rotational velocities, following the methodology described in \citet{Sacco2014A&A...565A.113S}. The spectral analysis of UVES spectra for F-, G-, and K-type stars—both in the Milky Way field and in stellar clusters, including GCs used for calibration—is thoroughly detailed in \citet{Smiljanic2014A&A...570A.122S} and \citet{Worley2024A&A...684A.148W}.

The final parameters and abundances are compiled in the {\sc dr5.1} catalogue, which includes the values used in this study: stellar parameters such as effective temperature \teff, surface gravity ($\log g$), metallicity ([Fe/H]), and the abundances of 32 elements, many in both neutral and ionised forms. For details about the line list, we refer the reader to \citet{Heiter2020}. This work focuses on the abundances of five elements predominantly produced by the $s$-process elements (Y, Zr, La, Ce, Ba) and three primarily generated by the $r$-process (Eu, Pr, Nd) in GCs. All abundances (by number) are given in logarithmic form: A(El) = 12 + $\log$(n(El)/n(H)), while abundances normalised to the solar scale \citep[from][]{Asplund_isoratios} are expressed as [El/H] = $\log$(n(El)/n(H)) -$\log$(n(El)/n(H))$_{\odot}$. 

\section{Trend with \vm}
\label{Sec:YBA_Vm}

As noted in the literature \citep[e.g.,][]{Worley2013}, strong absorption lines—such as those used to measure yttrium (Y) and barium (Ba)—are often saturated, making their derived abundances highly sensitive to the choice of microturbulence velocity (\vm). We refer the reader to Appendix \ref{App:vm}, to see the effect of \vm on each GCs. To reduce the impact of this dependence, we adopted the approach presented in \citetalias{Schiappacasse-Ulloa2024}, where a linear fit is applied to the original, uncorrected abundances from the {\it Gaia}-ESO catalogue (refereed from herafter as \textit{raw}) abundance versus \vm\ trend. The residuals from this fit, defined as $\Delta$[$El$/Fe], represent the corrected abundance values, where $El$ refers to a given element.

\begin{figure}
        \centering
        \includegraphics[width=0.9\columnwidth]{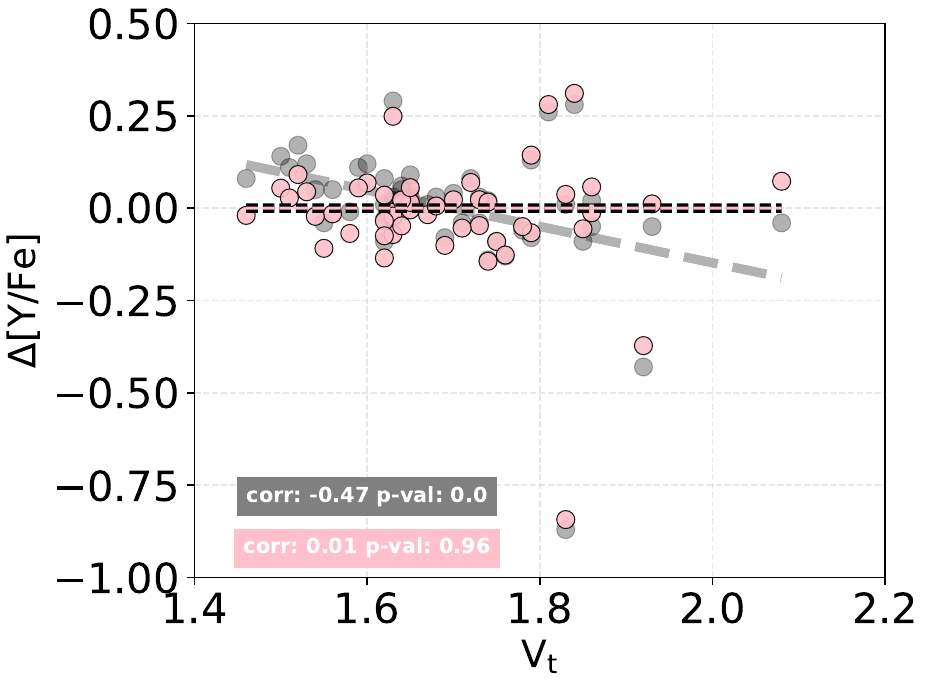}
        \caption{Example of the \textit{corrected} abundance of [Y/Fe] as a function of \vm for NGC~104. Grey and pink symbols show the \textit{raw} and \textit{corrected} abundance, respectively. See text for details.}
        \label{fig:ex_fit_Y}
\end{figure}

Figure~\ref{fig:ex_fit_Y} illustrates this method using yttrium in the GC NGC~104 as an example. Grey symbols show the raw [Y/Fe] abundances, while pink symbols represent the corrected values, $\Delta[Y/Fe]$. The figure also highlights the change in the Spearman correlation before and after the correction: the initial strong negative correlation with \vm\ (corr = -0.47 and p-val=0.00) becomes weak and statistically negligible (p-val = 0.96) after applying the correction. This correction method effectively removes artificial trends with \vm, which could otherwise introduce spurious internal scatter in the abundance distribution within a GC. Importantly, it preserves any genuine internal variations or correlations with other elements. However, because the method relies on an arbitrary zero-point, it does not provide absolute abundances. Therefore, the use of this method is only valid for internal comparisons within a given cluster.
For consistency, we applied this correction to all n-capture elements analysed in this study\footnote{The corrected abundances used in the present analysis are available in electronic format.}, including those with weaker lines (e.g., La and Eu), even though such elements are not expected to show significant sensitivity to \vm.

\section{Globular clusters in the Galactic context}
\label{Sec:Result_GalacticContext}

\subsection{Globular clusters as tracers of the Galaxy}

Among the interesting aspects of GCs is their ability to track the Galactic populations in a complementary way to the field stars \citep[see, e.g.][]{Massari2019, Massari2023A&A...680A..20M, Massari2025arXiv250201741M}. In addition, GCs offer a key advantage over field stars: their ages can be determined with higher precision through isochrone fitting of the entire cluster sequence \citep[see, e.g.][]{Chaboyer1995ApJ...444L...9C, Sarajedini1997PASP..109.1321S, VandenBerg2013ApJ...775..134V, Massari2023A&A...680A..20M, Aguado2025arXiv250220436A, Ceccarelli2025arXiv250302939C}. 

In this section, we aim to construct a large sample of GCs and field stars that covers the typical metallicity range of the Galactic halo, with precise abundances of n-capture elements. Our goal is to compare the behaviour of field stars and clusters, and to contrast these observations with the predictions of a chemical evolution model. In order to put both observation and models in the same scale, for this comparison, we used the raw average abundances of n-capture elements per cluster. To build our sample of field stars, we used two samples: the former is based on data from the Measuring at Intermediate Metallicity Neutron-Capture Elements (MINCE) project \citep{Cescutti2022}, which analysed high-quality spectra in a homogeneous and consistent manner, and the latter uses the results of  the Chemical Evolution of R-process Elements in Stars (CERES) project \citep{Lombardo2022}, including metal-poorer stars ([Fe/H]$<$–1.5). 
In particular, the MINCE results are outlined in: the first paper of the collaboration (MINCE I), in which the authors collected spectra of metal-poor stars ([Fe/H]$\lesssim$–1.5 dex), primarily from the Galactic halo; MINCE II \citep{Francois2024}, where they expanded their sample to include stars associated with two accreted remnants: Gaia-Sausage Enceladus (GSE) and Sequoia; most recently, MINCE III \citep{Lucertini2025}, in which they extended their study to stars with higher metallicities, up to [Fe/H]=–1.0 dex.

The abundance trends of our combined sample of field stars and GCs are compared with the results of a Galactic chemical evolution model obtained with the \texttt{GEMS} code \citep{rizzuti2025}, which is based on the stochastic models of \cite{2008A&A...481..691C} and \cite{Rizzuti2021}. 
We refer the reader to \citet{rizzuti2025} for a description of the model, and for some results for CNO and n-capture elements. Below is a brief overview of the key features and assumptions  of the model, while for  more details we refer to \citet{2015A&A...577A.139C, Rizzuti2019, Rizzuti2021, rizzuti2025}. Each of the many volumes in the simulations ($\sim$10,000) is considered isolated and contains the typical mass of gas swept by a SN-II explosion. The model can reproduce the evolution of the stars in the Galactic halo, with a total duration of 1 Gyr. For the nucleosynthesis of n-capture elements, the following sources are considered:  s-process from fast-rotating massive stars \citep{Limongi2018}; s-process from AGB stars \citep[\texttt{FRUITY,}][]{2011ApJS..197...17C}; r-process from neutron star mergers (NSMs) with 1 Myr time delay, with the Ba yields estimated by \cite{2013A&A...553A..51C} and the other elements scaled to the solar system r-process \citep{Cescutti2014}. The s-process by rotating massive stars is particularly important in this context, since rotation enables the production of s-process elements up to the second peak even at low metallicity. The question of the production sites of the  r-process is still open. However, several works have shown that the synthesis of Eu in the Galaxy can be explained by either only NSMs with a short time delay of 1 Myr, or both MRD SNe and NSMs with a longer time delay \citep[see, e.g][]{2014MNRAS.438.2177M, 2015A&A...577A.139C, Molero2023}.

\begin{figure*}
        \centering
        \includegraphics[width=0.95\textwidth]{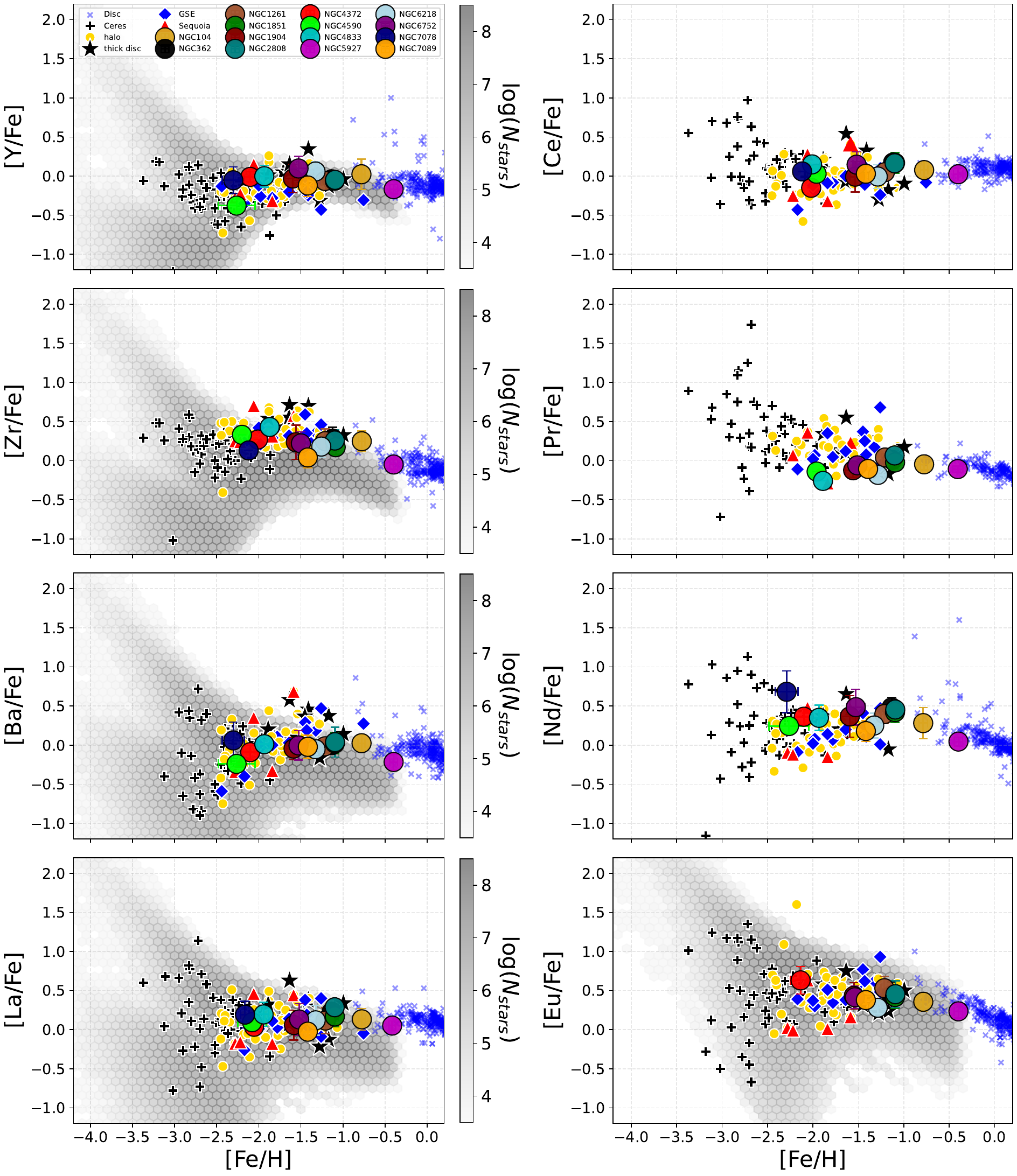}
        \caption{Comparison with Galactic chemical evolution models --when available-- with the reported [Y/Fe], [Zr/Fe], and [Ba/Fe], [La/Fe], [Ce/Fe], [Pr/Fe], [Nd/Fe], and [Eu/Fe]. The background, coloured in grey, scales with the logarithmic number of stars in each bin. We added in blue diamonds, red triangles, black pluses, yellow dots, blue crosses, and black stars the correspondent abundances of candidates stars of GSE \citep{Francois2024,Lucertini2025}, Sequoia \citep{Francois2024,Lucertini2025}, Ceres \citep{Lombardo2021,Lombardo2023}; halo stars \citep{Lucertini2025}, {\it Gaia}-ESO sample of disc stars, and thick disc stars \citep{Francois2024,Lucertini2025}, respectively.}
        \label{fig:ncap1_feh_all}
\end{figure*}

Figure \ref{fig:ncap1_feh_all} presents the raw average abundance ratios of Y, Zr, Ba, La, Ce, Pr, Nd, and Eu relative to iron in our GC sample (ed dots) as a function of metallicity [Fe/H], together with the sample of field stars from CERES and MINCE. We added also a sample of disc stars  from the {\it Gaia}-ESO survey. When available, model predictions are shown with a colour gradient representing the logarithmic number of stars in each bin.

First, both the GC and field star samples exhibit similar trends across all abundance planes, with an increasing spread towards lower metallicities.
When comparing model results to observations, it is notable that the stochastic models successfully reproduce both the dispersion arising from the contribution of different nucleosynthesis sites and the overall trend of chemical evolution. Additionally, model predictions suggest (super)solar abundances of [Y/Fe], [Zr/Fe], and [Ba/Fe] (as well as [La/Fe] and [Eu/Fe]) around [Fe/H] = -2.0 dex, followed by a decline at [Fe/H] > -1.5 dex, primarily due to contributions from Type Ia supernovae. The observed abundance trends in GCs closely match these predictions, particularly for the s-process elements Y, Ba, and La. A similar agreement is observed for the r-process element Eu. Interestingly, [Zr/Fe] aligns with the upper envelope of the model predictions, indicating a potential systematic offset, likely arising from the models' underestimation of the r-process contribution to Zr.

\subsection{The origin of globular clusters: hints from n-capture elements}

The population of Galactic GCs is known to have a diverse origin, with clusters broadly classified as either in-situ or ex-situ. We adopted the classification by \citet{Massari2019}, who combined kinematic and age information for a large sample of GCs to infer their likely origin. For methodological details, we refer the reader to the original publication. Table \ref{tab:in_ex_stats} summarises the origin of our sample based on this origin classification. As mentioned in \S\ref{Sec:YBA_Vm}, our corrected abundances ($\Delta$[$El$/Fe]) work only for internal comparisons, therefore, to ensure a consistent comparison between the two groups we used a slightly different approach. First, we grouped the GCs according to their origin, we filtered stars of each group to only those within a similar range of \vm (ranging from 1.35 and 2.35 km s$^{-1}$), excluding stars with different \vm ~values to minimise potential biases in the analysis. Then, we computed a \textit{global} $\Delta$ abundances across each group, defined as $\Delta$[$El$/Fe]$_g$. The table reports the number of selected stars, the maximum abundance difference (R$x$), and the interquartile range (IQR$x$) for each n-capture element, separated by in-situ and ex-situ GCs. In general, R$x$ and IQR$x$ values do not reveal significant differences between the two populations. The only exception is Eu, for which the IQR$x$ in in-situ GCs is 0.10 dex larger than in ex-situ clusters, suggesting a slightly broader internal variation.

\begin{figure*}
        \centering
        \begin{subfigure}
        \centering
        \includegraphics[width=0.65\columnwidth]{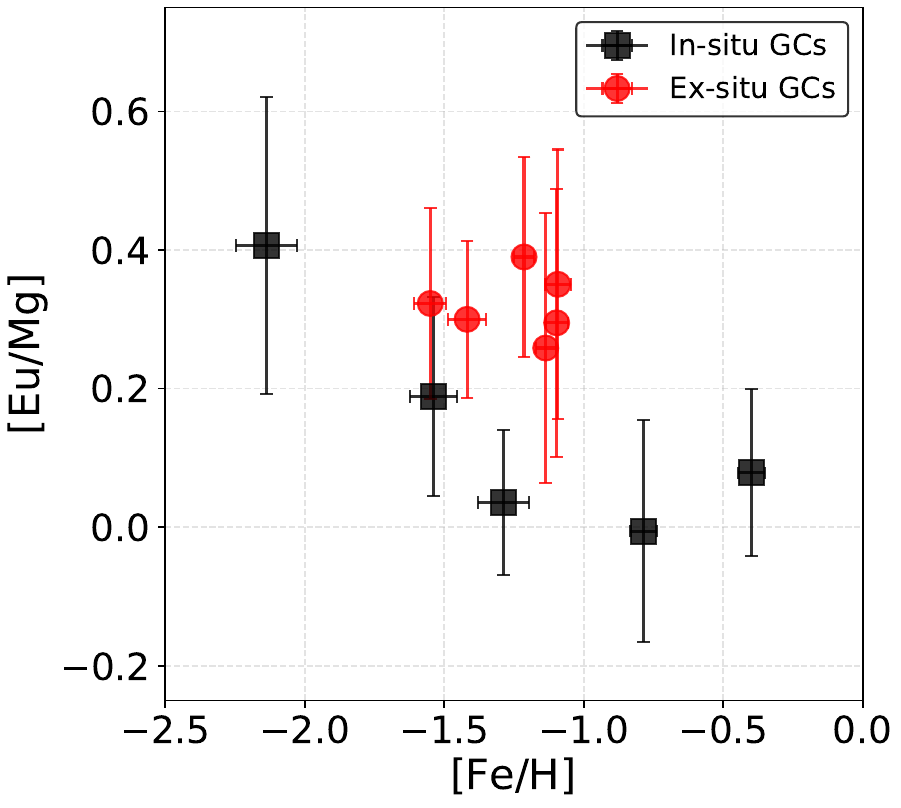}
        \end{subfigure}
        \hfill
        \begin{subfigure}
        \centering
        \includegraphics[width=0.6\columnwidth]{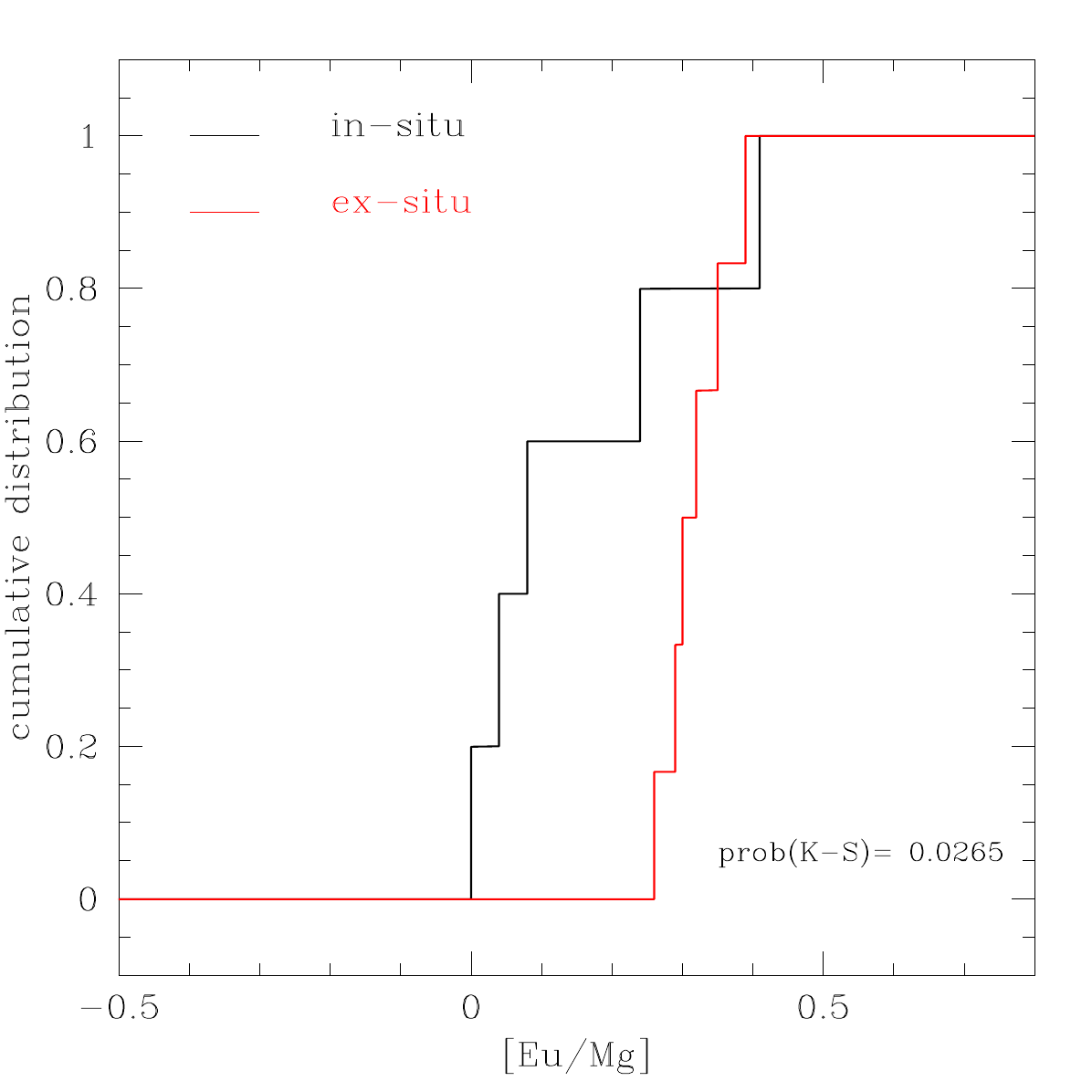}
        \end{subfigure}
        \hfill
        \begin{subfigure}
        \centering
        \includegraphics[width=0.635\columnwidth]{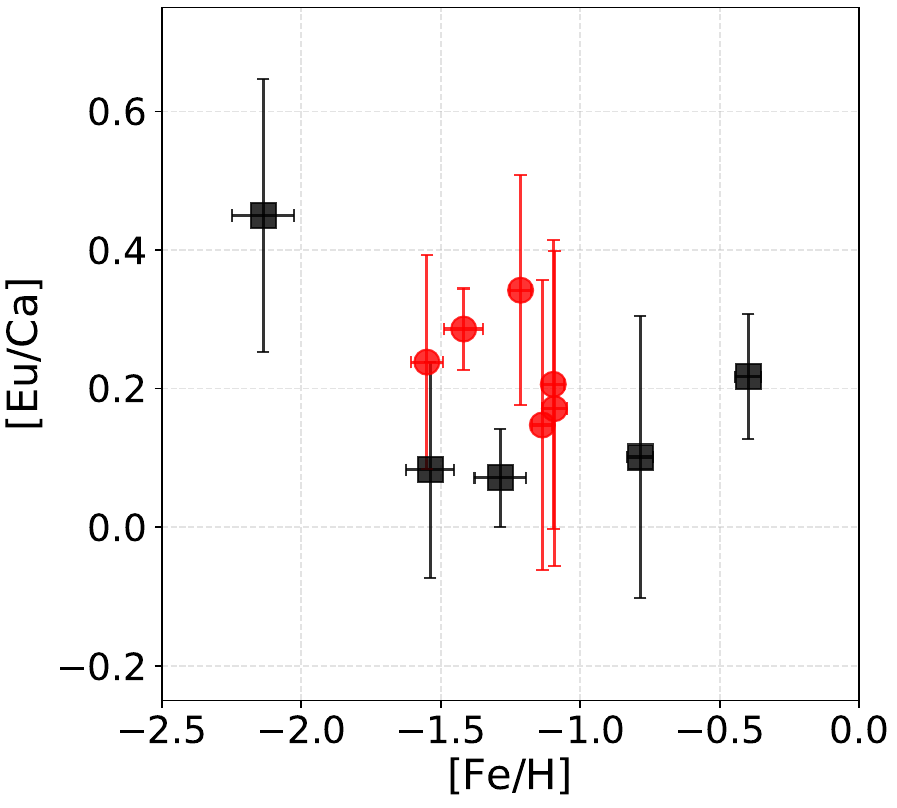}
        \end{subfigure}
        \centering
        \caption{The [Eu/Mg] and [Eu/Ca] ratios in the two sub-samples. While left panel displays the Eu over Mg ratio as a function of [Fe/H], the middle panel shows their (K-S) comparison of the cumulative distribution of the [Eu/Mg] ratio. The right panel display the Eu over Ca ratio as a function of [Fe/H].}
        \label{fig:EUMG_origin}
\end{figure*}

Some studies \citep[e.g.,][]{Monty2024} have suggested that combining r-process elements (e.g., Eu) with $\alpha$-elements (such as Mg or Si) may provide better insights into the origin of GCs. In particular, [Eu/Si] ratios in field stars with high and low orbital energy—typically associated with extra-galactic and Galactic origins, respectively—show distinct patterns. We tested this hypothesis in Figure \ref{fig:EUMG_origin}. The left panels display the raw [Eu/Mg] ratios versus [Fe/H] for our sample, grouped by origin according to \citet{Massari2019}. Note that [Eu/Mg] values are not available for a few clusters (NGC~4590, NGC~4833, and NGC~7078). In the figure, in-situ and ex-situ clusters are represented by black squares and red dots, respectively. While in-situ clusters exhibit a broad distribution of [Eu/Mg], ex-situ clusters show a flatter, systematically higher distribution at a given metallicity, with an average ratio of 0.14 and 0.32 dex, respectively. This offset likely reflects differences in the chemical enrichment histories of the two samples, consistent with the findings of \citet{Matsuno2021} and \citet{Monty2024}. Although one might suspect that the MP phenomenon—particularly Mg variations—could influence this ratio, a comparison with the results of \citet{Carretta2010} (see their Table 2) provides clarity. They reported the maximum [Mg/Fe] values found, corresponding to first-generation stars, for 19 GCs. On average, in-situ clusters show Mg abundances approximately 0.10 dex higher than their ex-situ counterparts. This supports the observed trend of lower [Eu/Mg] ratios among in-situ clusters. The right panel of Figure~\ref{fig:EUMG_origin} presents [Eu/Ca] ratios, where Ca serves as an $\alpha$-element unaffected by the MP phenomenon. Although the trend is more subtle, it shows a similar separation between the two populations, reinforcing the potential of [Eu/$\alpha$] ratios as tracers of GC origin. 
The middle panel of Fig.~\ref{fig:EUMG_origin} presents the Kolmogorov-Smirnov (K-S) comparison of the cumulative [Eu/Mg] distribution for in-situ (black line) and ex-situ (red line) GCs. This statistical test evaluates the likelihood that two samples originate from the same parent distribution. In this case, for our limited sample, a prob(K-S) of 0.027 allows us to reject the null hypothesis, reinforcing the conclusion that the [Eu/Mg] ratio effectively distinguishes ex-situ GCs.

To confirm in an independent way the chemical diversity of the two populations of GCs, we performed a simple clustering algorithm aiming to identify groups using only the chemical abundance features presented in the {\it Gaia}-ESO catalogue. To do so, we used a {\sc k-means} method \citep{Likas2003PatRe..36..451L} using the hyperparameter \textit{n\_cluster} equal to 2 in a subset of raw abundances ([Fe/H], [Na/Fe], [Mg/Fe], [Al/Fe], [Y/Fe], [Ba/Fe], [La/Fe], [Ce/Fe], [Eu/Fe], [Si/Fe], [Ca/Fe], [Sc/Fe], [Ti/Fe], [V/Fe], [Cr/Fe], [Mn/Fe], [Co/Fe], [Ni/Fe], [Cu/Fe],and [Zn/Fe]). 
The results showed that, for all the GCs with available Eu abundances, the clustering algorithm was able to correctly identify the origin label (as provided by \citep{Massari2019}) in 10 out of 11 cases, failing only for NGC~1904, for which it determined an in-situ origin. This independently confirms that they are chemically distinct.

\begin{table}[!ht]
    \centering
   \caption{Average and IQR for all the elements analysed for GCs with different origin. The listed the total number of stars (N) considered for the statistics.}
   \label{tab:in_ex_stats}
    \begin{tabular}{c|ccc|ccc}
    \hline
    \hline
        & \multicolumn{3}{c|}{In-Situ GCs} & \multicolumn{3}{c}{ex-situ GCs}     \\
        ~ & N & R(x) & IQR(x) & N & R(x) & IQR(x) \\ \hline
        $\Delta${[}Y/Fe{]}$_g$  & 106 &  0.76 & 0.13 & 114 & 0.85 & 0.14 \\ 
        $\Delta${[}Zr/Fe{]}$_g$ & 67  &  0.71 & 0.25 & 65  & 0.70 & 0.20 \\ 
        $\Delta${[}Ba/Fe{]}$_g$ & 104 &  0.64 & 0.10 & 112 & 0.71 & 0.10 \\ 
        $\Delta${[}La/Fe{]}$_g$ & 67  &  0.71 & 0.19 & 73  & 0.74 & 0.21 \\ 
        $\Delta${[}Ce/Fe{]}$_g$ & 51  &  0.67 & 0.11 & 51  & 0.67 & 0.15 \\ 
        $\Delta${[}Pr/Fe{]}$_g$ & 48  &  0.52 & 0.14 & 48  & 0.35 & 0.09 \\ 
        $\Delta${[}Nd/Fe{]}$_g$ & 105 &  0.83 & 0.21 & 109 & 0.83 & 0.18 \\ 
        $\Delta${[}Eu/Fe{]}$_g$ & 80  &  0.65 & 0.31 & 78  & 0.71 & 0.20 \\ 
    \hline
    \end{tabular}
\end{table}

\section{N-capture abundances and other chemical multi population indicators}
\label{Sec:n-cap_vs_NAL}

The two elements most commonly used to identify the different stellar populations in GCs are Na and Al. They are produced during hot H-burning and thus are involved in the MP phenomenon. In this analysis, we classify FG stars as those within three times the typical observational Na uncertainty (0.075 dex) from [Na/Fe]$_{lim}$. The latter is defined as the raw average [Na/Fe] abundance of the three stars with the lowest [Na/Fe] values. The following paragraphs present our findings on the correlations between light and n-capture element abundances across different stellar populations in GCs.

The presence of possible correlations between the abundances of Na and Al with those of the s-process elements may give us clues as to the nature of polluters that changed the composition of SG stars. Indeed, both Na and Al and s-process elements can be produced in similar environments, particularly in intermediate- to low-mass stars \citep[see e.g.,][]{Karakas2014} or in massive stars with rotation-induced mixing \citep[see e.g.,][]{Limongi2018}. 
Therefore, a shared origin for both elements in specific types of stars might lead to a correlation between them. On the other hand, we do not expect correlations between the abundances of Na and Al with those of the r-process elements.

For this study, we assessed the significance of correlations using the p-value from the Spearman statistical method \citep{Spearman1904}. We classified correlations as highly significant for p-values below 0.05, moderately significant for p-values between 0.05 and 0.10, and weak or negligible for p-values greater than 0.10. It is important to note that outlier abundances --which were defined as abundances larger or lower than 0.5 dex than the average cluster abundance-- were excluded from this analysis.

\begin{figure*}
        \centering
        \includegraphics[width=0.9\textwidth]{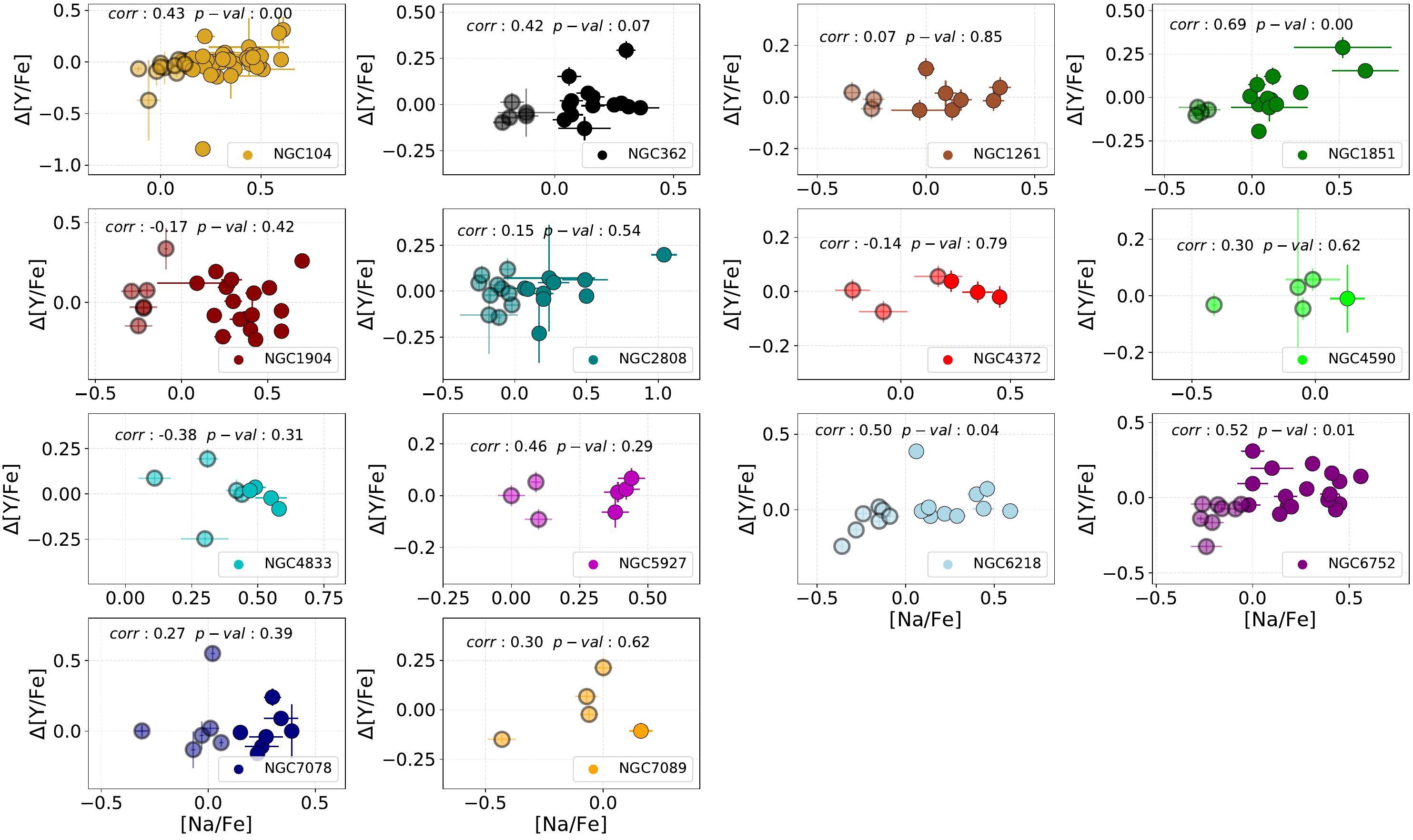}
        \caption{$\Delta$[Y/Fe] abundances as a function of [Na/Fe] for GCs in our sample. Lighter and darker symbols represent stars belonging to the FG and SG according to our definition (see text), respectively. Every panel shows the correspondent Spearman correlation and p-value.}
        \label{fig:corr_Y_Na}
\end{figure*}

\paragraph{Sodium:} Fig.\ref{fig:corr_Y_Na} shows the abundance of Y as a function of the Na abundances for our sample of GCs. We found mixed results for well-sampled GCs. While some clusters — such as NGC~104, NGC~362, NGC~1851, NGC~6218, and NGC~6752 — exhibited mildly or highly significant correlations, others showed no statistically significant trends. The correlation factors are consistently positive. Figures \ref{fig:corr_na1} and \ref{fig:corr_na2} present the corresponding results for all other analysed n-capture elements, including Zr, Ba, La, Ce, Pr, Nd, and Eu. 
The other first-peak s-process element, Zr, exhibits a mildly significant positive correlation in two globular clusters (NGC~104 and NGC~1851). Among the second-peak s-process elements, Ba shows a highly significant positive correlation in NGC~362, and NGC~6218, whereas La does not display any significant correlation in any cluster.
Ce exhibits two distinct behaviors: in NGC~6218, it shows a highly significant positive correlation, while in other clusters, its trend is less clear. In contrast, Pr shows no significant correlation. Nd, on the other hand, presents a highly significant positive correlation in NGC~1851 and NGC~6218, and a mildly significant one in NGC~362 and NGC~6752.
Finally, the r-process element Eu does not exhibit any correlation in any of the studied globular clusters.
    
\paragraph{Aluminum:} Analogously, we compare the abundances of n-capture elements with those of Al. While Fig.\ref{fig:corr_al1} show the $\Delta$[Y/Fe], $\Delta$[Zr/Fe], $\Delta$[Ba/Fe], and $\Delta$[La/Fe] abundances, Fig. \ref{fig:corr_al2} displays the $\Delta$[Ce/Fe], $\Delta$[Pr/Fe], $\Delta$[Nd/Fe], and $\Delta$[Eu/Fe] abundances as a function of the [Al/Fe] for our sample of GCs. Although in many clusters we did not observe any significant correlation between n-capture element abundances and those of Al, for the case of $\Delta$[Y/Fe], the GCs NGC~104, NGC~1851, NGC~5927, NGC~6218, NGC~6752, and NGC~7089 show a highly significant positive correlation, while NGC~362 and NGC~2808 exhibit a mild correlation. Zr only presents high significant positive correlations in NGC~1261, and a mild one in NGC~104; Ba shows a highly significant positive correlation in NGC~362 and NGC~6218 and a mild one in NGC~6752. La only has a highly significant positive correlation in NGC~104. Ce presents different behaviours, with NGC~104 having a mildly significant positive correlation, a highly significant one in NGC~6218 and a negative highly significant one in NGC~1904. A similar case for Pr presents a highly significant positive correlation in NGC~104 but a negative one with mild significance in NGC~1904. Nd shows a positive correlation in NGC~104, NGC~1851, NGC~6218, NGC~7078, NGC~7089, and NGC~1261 with high significance in the first five clusters and mild one in the last one. Finally, Eu exhibits a highly significant negative correlation in NGC~5927 and a positive one in NGC~7089.

Although we found many (anti-)correlations among different pairs of elements, their strengths vary cluster-to-cluster. This means that the Na/Al nucleosynthetic production affects the Y/Zr/Ba/La/Ce/Pr/Nd/Eu nucleosynthesis in those GCs in a positive or negative way, depending on the slope of the correlation. We can conclude that, in general, correlations with slow n-capture elements show some associations with Na and Al while r-process elements like Eu do not. This seems to suggest that slow n-capture elements and light elements like Na and Al could share common nucleosynthesis sites, supporting the hypothesis that AGB stars of different masses and/or massive stars with rotation-induced mixing could contribute to the intra-cluster pollution. On the other hand, the significance and magnitude of these correlations are not always consistent, nor are explainable in nucleosynthetic terms (e.g. La and Ce production is closely associated to that of Ba, therefore the correlation of one of these elements with Al or Na without a similar behaviour from the other two is of limited astrophysical significance). 
This may be attributed to limited sample sizes, variations in behavior across different metallicities, and potentially unaccounted-for observational uncertainties. Similar results have been reported in the literature; in particular, \citet{Yong2008} found correlations between some light s-process elements and light elements in NGC~1851.

\begin{figure}
        \centering
        \includegraphics[width=\columnwidth]{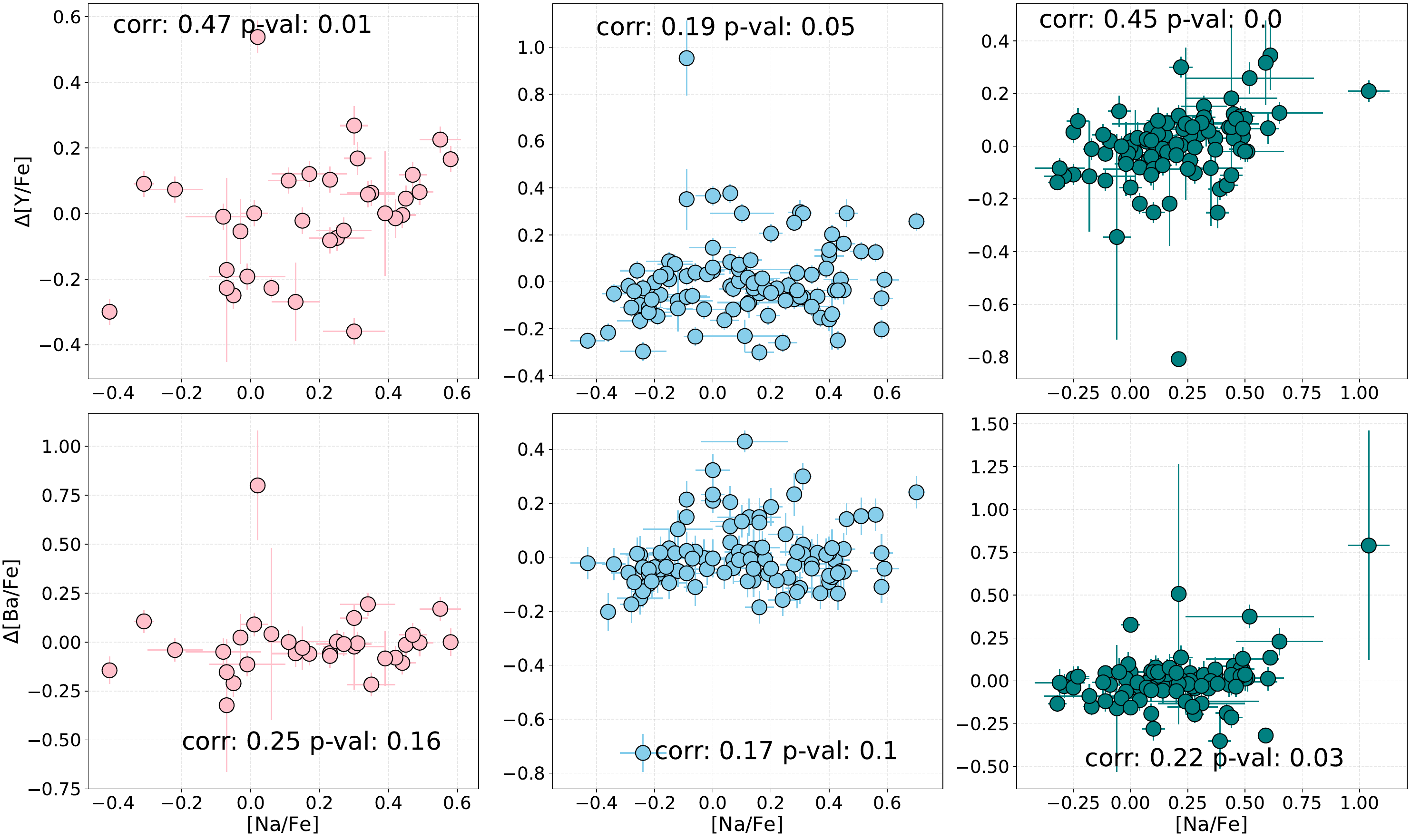}
        \caption{Distribution of the $\Delta$[Y/Fe] (upper panels) and $\Delta$[Ba/Fe] (lower panels) as a function of [Na/Fe] divided by metallicity bins. The metallicity bins from left to right are: [Fe/H]<-1.8, -1.8<[Fe/H]<-1.1, and [Fe/H]>-1.1 dex.}
        \label{fig:YBa_Na}
\end{figure}

To improve our statistical robustness, we focused on global trends across GCs rather than analysing them individually, following the methodology outlined in \citetalias{Schiappacasse-Ulloa2024}, who explored potential correlations between s-process elements and Na by dividing their sample into three metallicity regimes: metal-poor, mid-metallicity, and metal-rich for which $\Delta$s were determined as a group. These regimes correspond to GCs with metallicities of [Fe/H] < -1.8, -1.8 < [Fe/H] < -1.1, and [Fe/H] > -1.1 dex, respectively. In their study, they reported a slight correlation between Y and Na. To test this in our sample, we applied the same procedure. Figure \ref{fig:YBa_Na} presents the results for our sample along with the corresponding Spearman correlation coefficients, calculated without including outlier stars. The upper panels display the distribution of $\Delta$[Y/Fe] as a function of [Na/Fe], while the lower panels show the analogous results for $\Delta$[Ba/Fe]. Each panel includes the correlation coefficient and its associated significance, which reflect a strong correlation between $\Delta$[Y/Fe] and [Na/Fe] in all the metallicity regimes. On the other hand, $\Delta$[Ba/Fe] and [Na/Fe] display a significant correlation only in the most metal-rich regime, which disappears while decreasing the metallicity. Our results support those of \citetalias{Schiappacasse-Ulloa2024}, confirming a significant contribution of light s-process elements in the same nucleosynthetic site where Na was produced.

\section{Neutron-capture homogeneity in GCs}
\label{Sec:n-cap_GCs}

The degree of chemical homogeneity within GCs offers valuable clues about their origin. Elements such as Al, Na, and O are known to exhibit significant internal spreads \citep{Carretta2009u}, while most heavier elements remain relatively uniform across the majority of GCs—commonly referred to as Type I clusters. In contrast, Type II GCs display more complex abundance patterns, including variations in iron and, in some cases, s-process elements. Based on classifications by \citet{Milone2017} and \citet{Nardiello2018}, our sample includes five Type II clusters: NGC~362, NGC~1261, NGC~1851, NGC~7078, and NGC~7089. The remaining clusters are classified as Type I, except NGC~1904 and NGC~4372, which lack definitive classifications. Despite growing interest, evidence of internal variations in n-capture elements within GCs remains limited, with only a few studies—such as the recent analysis of NGC~6341 by \citet{Kirby2023}—reporting such findings.

\begin{figure*}
        \centering
        \begin{subfigure}
        \centering
        \includegraphics[width=0.90\columnwidth]{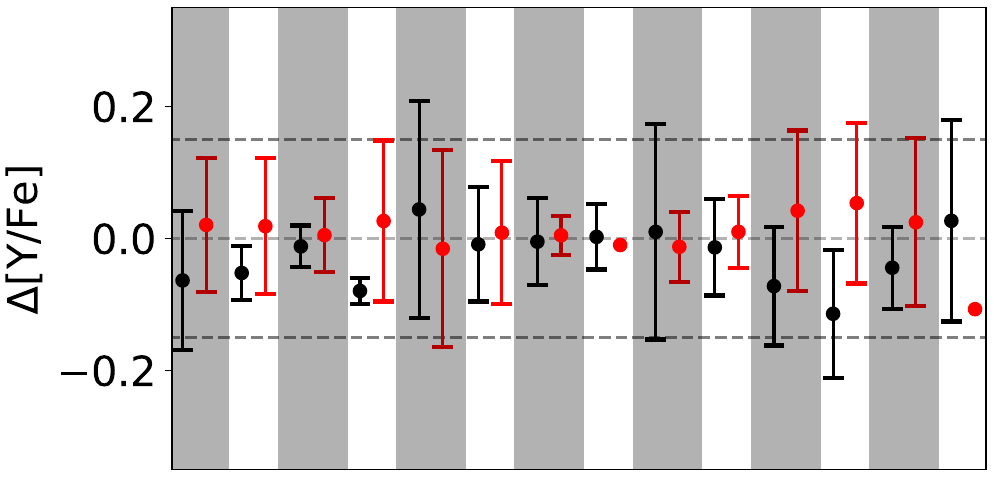}
        \end{subfigure}
        \hfill
        \begin{subfigure}
        \centering
        \includegraphics[width=0.90\columnwidth]{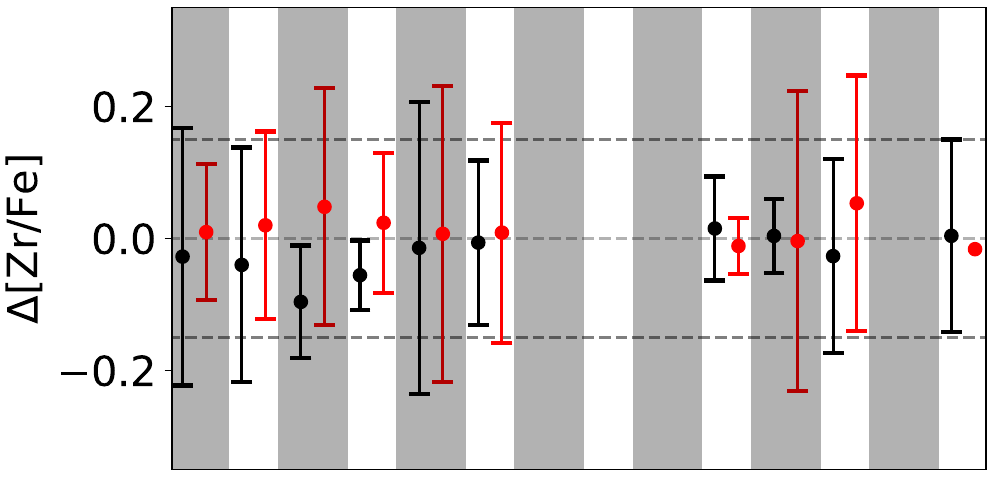}
        \end{subfigure}
        \vfill
        \begin{subfigure}
        \centering
        \includegraphics[width=0.90\columnwidth]{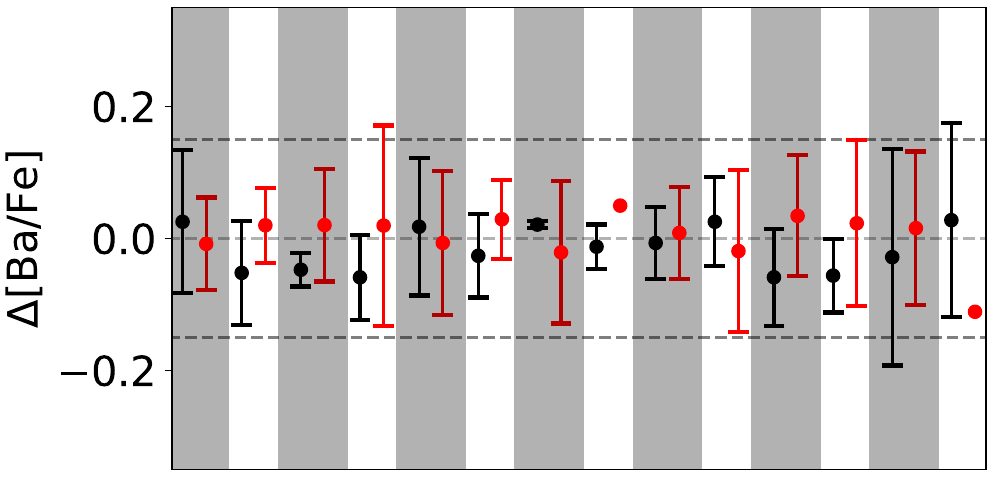}
        \end{subfigure}
        \hfill
        \begin{subfigure}
        \centering
        \includegraphics[width=0.90\columnwidth]{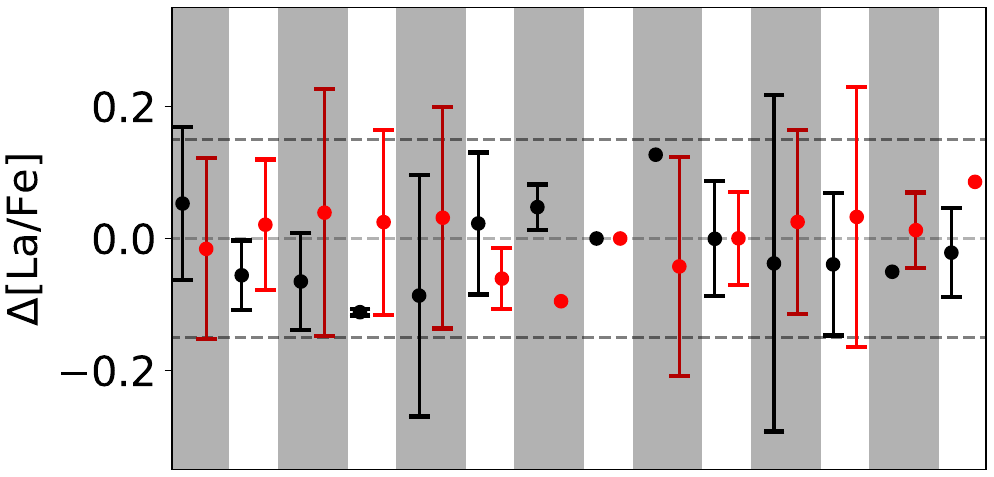}
        \end{subfigure}
        \vfill
        \begin{subfigure}
        \centering
        \includegraphics[width=0.90\columnwidth]{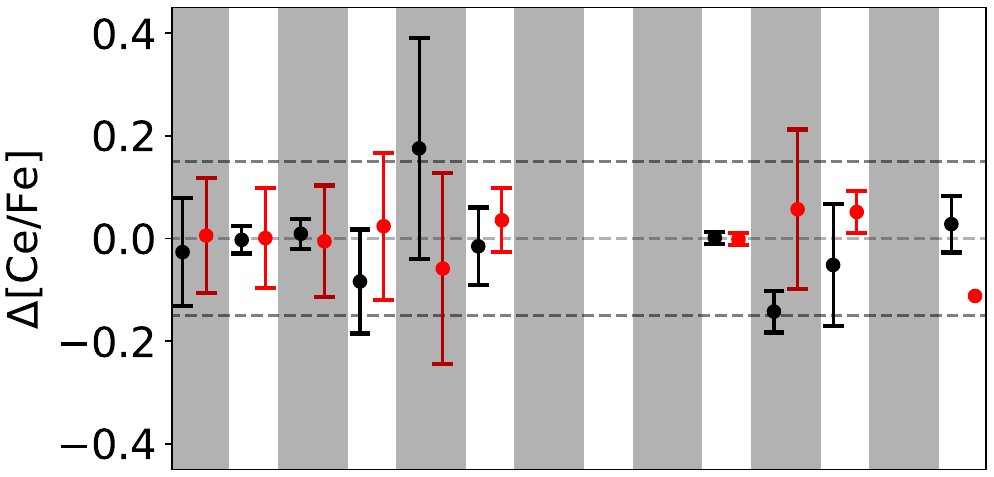}
        \end{subfigure}
        \hfill
        \begin{subfigure}
        \centering
        \includegraphics[width=0.90\columnwidth]{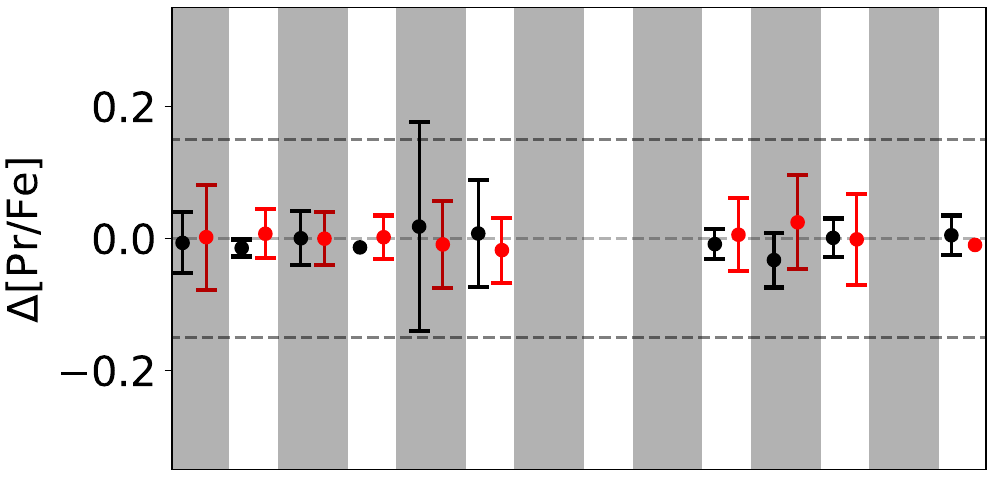}
        \end{subfigure}
        \vfill
        \begin{subfigure}
        \centering
        \includegraphics[width=0.90\columnwidth]{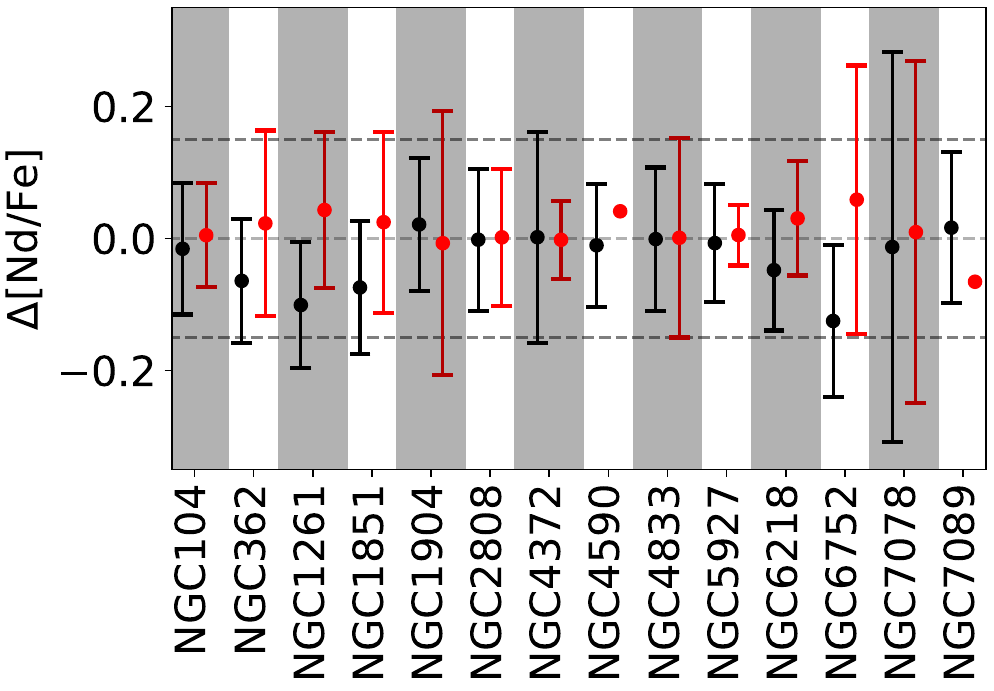}
        \end{subfigure}
        \hfill
        \begin{subfigure}
        \centering
        \includegraphics[width=0.90\columnwidth]{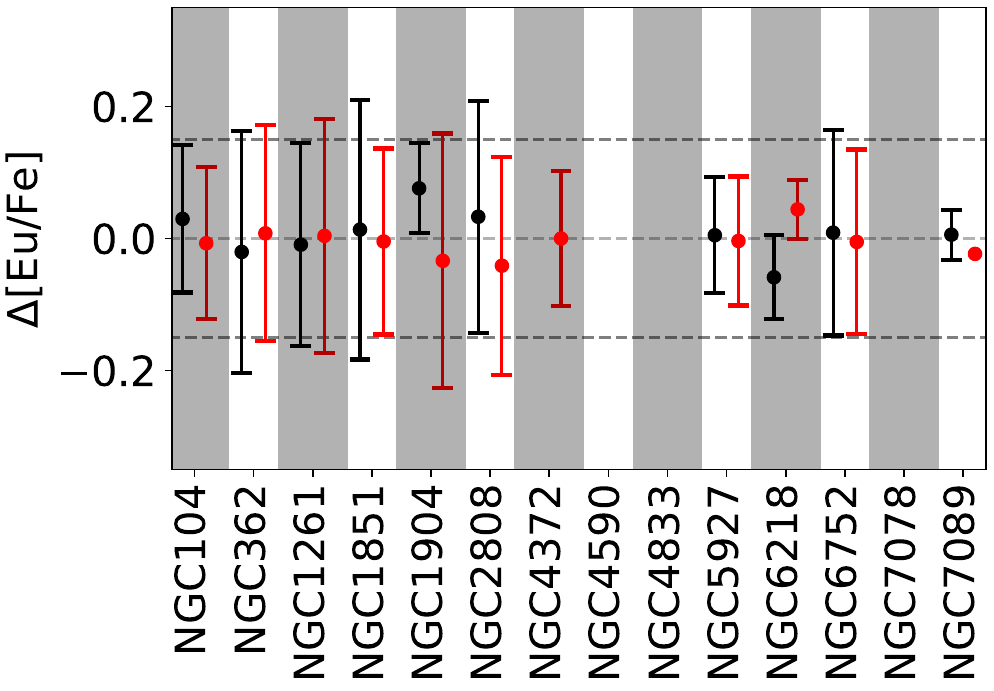}
        \end{subfigure}
        \hfill
        \caption{Average and standard deviation for $\Delta$[$El$/Fe] per GCs. They are grouped by generation, with black and red symbols FG and SG stars, respectively. White and grey areas are plotted only to facilitate the figure's readability and do not have further meaning. Dashed lines at $\pm$0.15 dex represent the typical uncertainty in the abundance's measurements.}
        \label{fig:spread_gen}
\end{figure*}

We examined whether the internal abundance spreads differ between the two stellar populations within each GC. To separate the FG and SG stars, we applied the classification described in Section~\ref{Sec:n-cap_vs_NAL}. Figure~\ref{fig:spread_gen} displays the average and standard deviation of $\Delta$[$El$/Fe] for FG (black symbols) and SG (red symbols) stars in each GC and for each element. Dashed lines at $\pm$0.15 dex represent a conservative typical error in the measurement of the abundances. Metal-poor clusters such as NGC~4372, NGC~4590, NGC~4833, NGC~7078, and NGC~7089 lack sufficient data to support a robust comparison. While we observed differences in the average abundances between FG and SG populations, these differences typically fall within one standard deviation of each other. This may be due to the division of the sample into subpopulations, which reduces the number of stars in each group and increases statistical uncertainty. Consequently, genuine underlying trends may appear less significant, even if a broader correlation is present.

In most Type I GCs, both populations show comparable spreads. In fact, the ratio of the spreads between the second- and first-generation is about 1, suggesting that n-capture elements are minimally affected by the polluters responsible for the MP phenomenon. For Type II GCs --specifically in GCs with larger number statistics (NGC~362, NGC~1261, and NGC~1851)-- the SG stars consistently show larger spreads in s-process elements, having on average a spread ratio between the two populations generation of about twice larger than Type I GCs, indicating the contribution from polluters to the chemical enrichment of the clusters. These finding supports previous study that reported spread in s-process elements in NGC~362 \citep[e.g.,][]{Monty2023} and NGC~1851 \citep{Carretta2011}. In the case of NGC~1261, neither \citet{Koch-Hansen2021} nor \citet{Marino2021} found significant spreads in s-process elements. Nevertheless, \citet{Koch-Hansen2021} reported a peculiar overabundance of r-process elements, which may also contribute, to some extent, to the production of s-process elements.

Further analysis using differential analysis techniques \citep[e.g.,][]{Monty2023} has been shown to be effective in overcoming current measurement uncertainties, contributing to a more precise determination of the internal variation in GCs. In the longer term, high-resolution, multi-object spectrographs will be essential to fully explore these subtle chemical variations. Instruments like the proposed HRMOS \citep{Magrini2023} will enable precise abundance determinations for a wide range of n-capture elements in GC stars -including elements of the third peak, like Pb-offering deeper insights into the nature and contributions of intra-cluster polluters.

\section{Summary and Conclusions}
\label{Sec: Summary}

We utilised the GC sample of the {\it Gaia}-ESO survey to investigate the abundances of n-capture elements. First, we analysed the abundances of Y, Zr, Ba, La, Ce, Nd, Pr, and Eu to trace the chemical composition of the Galaxy. 
We combined our results with those of field stars (MINCE and CERES samples) and compared with the results of a stochastic chemical evolution model of the Milky Way halo \citep{rizzuti2025}. 
Then, we aimed to differentiate the origins of these clusters by examining [Eu/Mg], a ratio that reflects the distinct chemical enrichment histories of their birth environments. 
Next, we explored potential relationships between hot H-burning elements—commonly used as tracers of MPs in GCs—and heavier elements to gain insights into the nature of the stars responsible for intracluster pollution. Finally, we investigated the homogeneity of GCs in n-capture abundances, considering also separately their first and second generation stars. Below, we summarize our key findings:

(i) Most GCs in our sample closely follow the chemical distribution of halo and disc field stars with similar metallicity across all analysed n-capture elements. 
In addition, model predictions generally align well with the observed trends for most elements. The main exception is Zr, where observations cover the upper envelope of the model predictions, likely due to an underestimation of the r-process contribution to Zr nucleosynthesis. 

(ii) The [Eu/Mg] ratio appears to be a useful indicator of GC origin—at least in distinguishing in-situ from ex-situ clusters—reflecting the different chemical enrichment histories of their formation environments. This finding reinforces previous results in the literature. The classification of GCs based on their chemical properties was further supported by a clustering algorithm, which successfully recovered the origin labels for most of the sample where Eu measurements were available. However, a larger GC sample covering a wider metallicity range is needed to confirm this trend.

(iii) We explored n-capture element abundances in relation to Na and Al, which are typically used as tracers of MPs in GCs. Some clusters, such as NGC~104, NGC~1851, NGC~6218, and NGC~6752, displayed strong correlations between the light s-process element Y and Na. Similarly, clusters like NGC~362, and NGC~6218 showed significant correlations between the heavier s-process element Ba and Na. However, no correlations were found with the r-process element Eu.
When dividing the GC sample into different metallicity bins—low-metallicity ([Fe/H] < -1.8), intermediate-metallicity (-1.8 < [Fe/H] < -1.1), and high-metallicity ([Fe/H] > -1.1 dex)—we observed a strong correlation between Y and Na in all three regimes. These results suggest that light s-process elements might have been synthesised in the same nucleosynthetic sites as Na, with the strength of their contribution varying based on the mass and metallicity of the polluter, like AGB stars and/or fast-rotating massive stars.

(iv) Finally, we found no strong evidence of systematic differences in abundance spreads between FG and SG stars in Type I GCs. In contrast, in Type II GCs with large sample sizes—such as NGC~362, NGC~1261, and NGC~1851—SG stars consistently exhibited larger spreads in s-process element abundances compared to FG stars, suggesting a possible influence from internal polluters.

It is important to note that internal abundance spreads in GCs are often subtle, making them difficult to detect given observational uncertainties. Therefore, we emphasise the need for a high-resolution multi-object spectrograph, such as HRMOS, to overcome these limitations and provide more precise measurements.

\begin{acknowledgements}
J.S.U.,  L.M., S.R., G.C., F.R. and L.B. thank INAF for the support (Large Grant EPOCH), the Mini-Grants Checs (1.05.23.04.02), and the financial support under the National Recovery and Resilience Plan (NRRP), Mission 4, Component 2, Investment 1.1, Call for tender No. 104 published on 2.2.2022 by the Italian Ministry of University and Research (MUR), funded by the European Union – NextGenerationEU – Project ‘Cosmic POT’ Grant Assignment Decree No. 2022X4TM3H by the Italian Ministry of the University and Research (MUR). FR is a fellow of the Alexander von Humboldt Foundation. FR acknowledges support by the Klaus Tschira Foundation. Based on data products from observations made with ESO Telescopes at the La Silla Paranal Observatory under programmes 188.B-3002, 193.B-0936, and 197.B-1074. The {\it Gaia}-ESO final data release {\sc dr5.1} is available from the \hyperlink{https://www.eso.org/qi/catalog/show/411}{ESO webpage}.
\end{acknowledgements}

%
%

\bibliographystyle{aa}
\bibliography{bibliography}

\begin{appendix}

\section{Comparison with the literature}
\label{Sec:comp_lit}
We compared the results of {\it Gaia}-ESO {\sc dr5.1} with the literature ones. To this aim, we used the results of the spectroscopic analysis done by \citet{Carretta2009u} (hereinafter \citetalias{Carretta2009u}), who analysed more than 200 stars in 18 GCs using UVES spectra. For these spectra, they obtained stellar parameters and abundances of Na, Mg, Al, and Si. Moreover, \citetalias{Schiappacasse-Ulloa2024} extended the analysis to n-capture elements using the same sample and adopting their stellar parameters. In the mentioned sample, we have in common with our {\it Gaia}-ESO  sample 48 stars in six GCs. Here we compared their stellar parameters as well as the abundances of Na, Mg, Al, Si from \citetalias{Carretta2009u} and  Y, Ba, La, and Eu from \citetalias{Schiappacasse-Ulloa2024}. In addition, we added the GCs NGC~1851 \citep{Carretta2011}, NGC~362 \citep{Carretta2013}, and NGC~4833 \citep{Carretta2014}, which were analysed in analogous way as the aforementioned articles. In the latter set of GCs, we have 37 stars in common, for which the stellar parameters and the abundance of Na, Mg, Al, Si, Y, Ba, La, Eu were measured.

\begin{figure*}
        \centering
        \includegraphics[width=\textwidth]{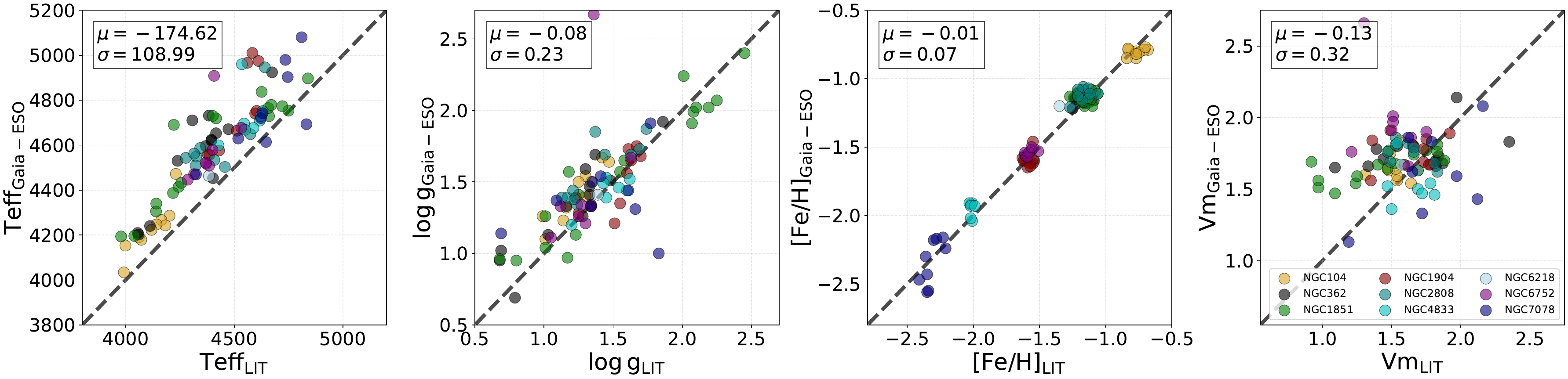}
        \caption{Comparison of the stellar parameters reported by \citetalias{Carretta2009u}, \citet{Carretta2011}, \citet{Carretta2013}, \citet{Carretta2014}, \citetalias{Schiappacasse-Ulloa2024}, and the {\it Gaia}-ESO {\sc dr5.1} for in-common stars. Each panel shows the mean and standard deviation of the difference between the result obtained by {\it Gaia}-ESO and \citetalias{Carretta2009u}.}
        \label{fig:ste_par_carr}
\end{figure*}

Fig.\ref{fig:ste_par_carr} shows the \teff, $\log$ g, [Fe/H], and microturbulence velocity (\vm) reported by {\it Gaia}-ESO {\sc dr5.1} and their corresponding results obtained by \citetalias{Carretta2009u}, \citet{Carretta2011}, \citet{Carretta2013}, and \citet{Carretta2014}. Each panel displays the stars in common, colour-coded by GC, along with the mean ($\mu$) and standard deviation ($\sigma$) of the differences between the literature and {\it Gaia}-ESO values. The panels show that there is a systematic difference toward higher \teff\, ~and $\log$ g in the results derived by {\it Gaia}-ESO. It is worth noticing that the \teff~ reported by \citetalias{Carretta2009u} were obtained using photometric relations, and the $\log g$ were obtained using the mentioned \teff\ along with the corresponding bolometric corrections. Therefore, those systematic differences could be due to the different methods used for their determination. 

There is no clear correlation between the two \vm determinations because both cover a small range and remain nearly constant. This is due to the sample being almost entirely composed of red giant stars with similar \vm values. However, on average, the \vm values of {\it Gaia}-ESO are 0.13 higher than the ones reported in \citetalias{Carretta2009u}, which is about their typical internal error found in the \vm determination. 

\begin{figure*}
        \centering
        \includegraphics[width=\textwidth]{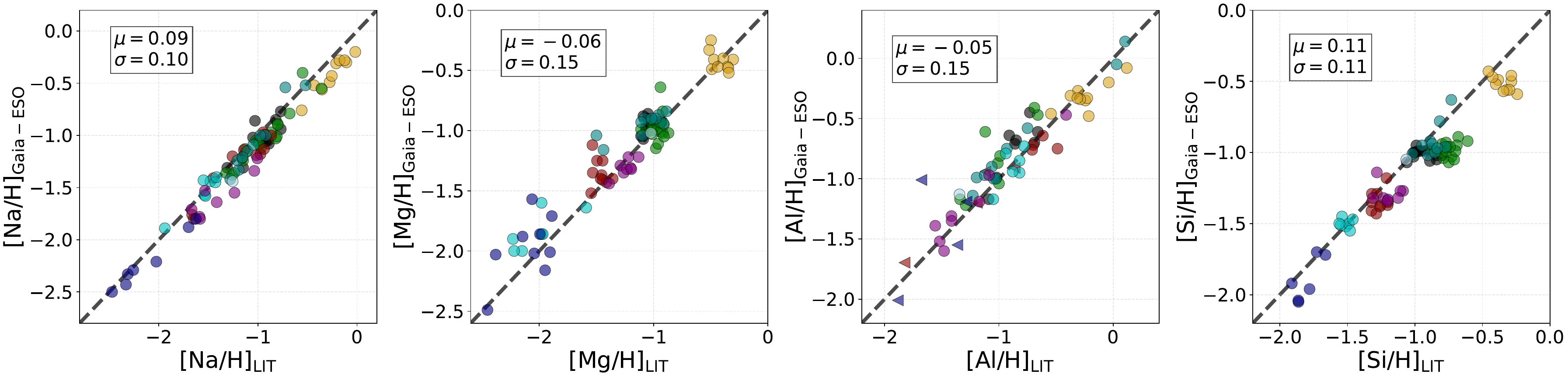}
        \caption{Comparison of the (from left to right) [Na/H], [Mg/H], [Al/H], and [Si/H] reported by \citetalias{Carretta2009u}, \citet{Carretta2011}, \citet{Carretta2013}, \citet{Carretta2014}, \citetalias{Schiappacasse-Ulloa2024}, and the {\it Gaia}-ESO {\sc dr5.1} for in-common stars. Coloured dots and triangles represent actual measurements and upper limits, respectively. The present figure follows the same description as Fig.\ref{fig:ste_par_carr}.}
        \label{fig:abund_carr}
\end{figure*}

Fig. \ref{fig:abund_carr} displays the comparison of [Na/H], [Mg/H], [Al/H], and [Si/H] normalised to the Sun for the in-common stars. On average, the {\it Gaia}-ESO {\sc dr5.1} results agree with the literature. It is worth noticing that upper limits (triangle symbols) reported by \citetalias{Carretta2009u} were not used for the computation of the mean and standard deviation of the results' difference. In general, all the species follow closely the one-to-one relation. [Na/H] and [Si/H] from {\it Gaia}-ESO show systematically lower abundance which could be due to the higher \teff ~and $\log$ g\footnote{We ensured to report both sets of abundances on the same solar scale}. The [Mg/H] and [Al/H] abundances of {\it Gaia}-ESO are slightly higher in comparison with the literature, which might be due to  the slightly higher \teff computed by the survey.

\begin{figure*}
        \centering
        \includegraphics[width=\textwidth]{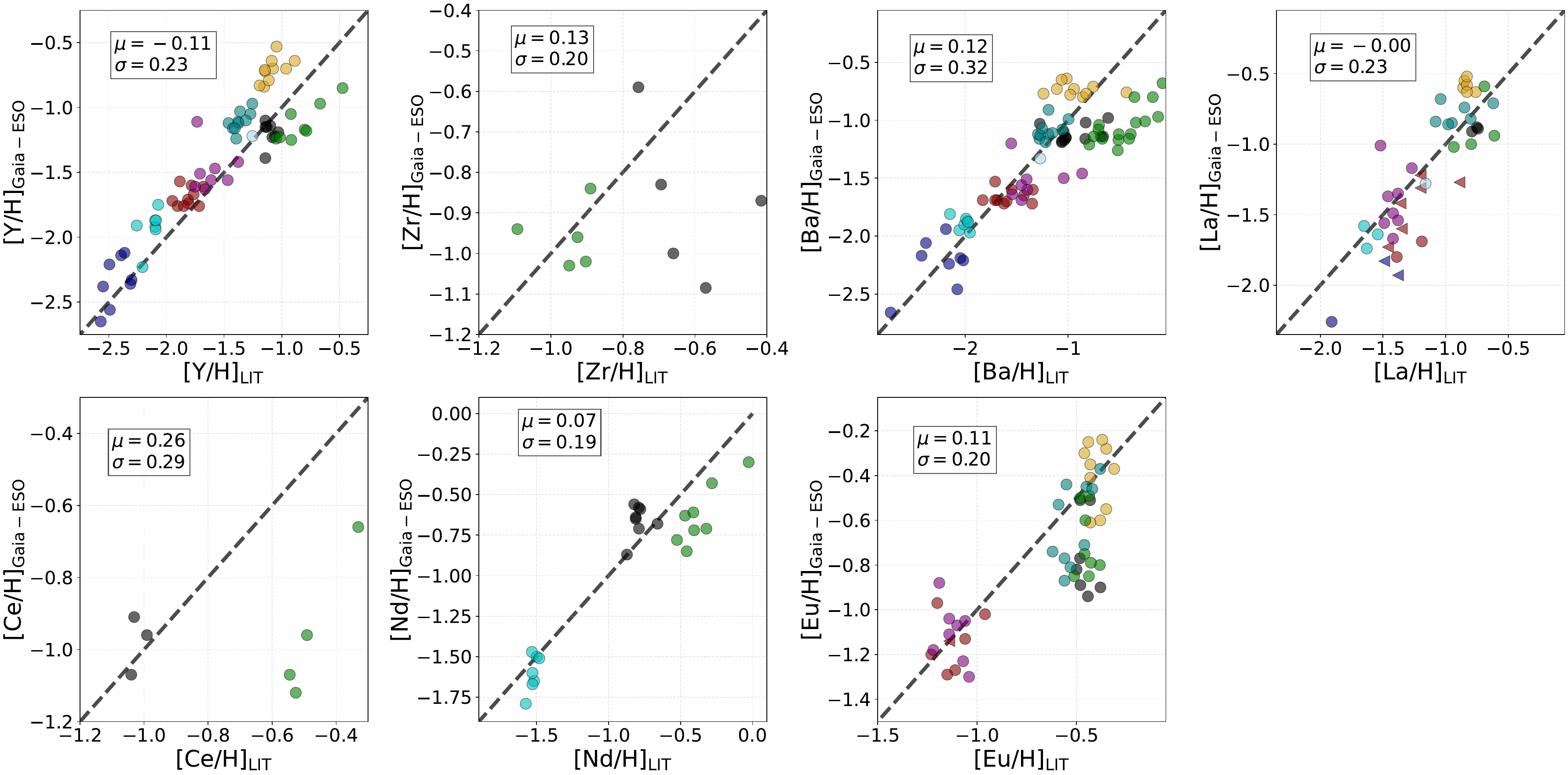}
        \caption{Comparison of the abundance of (from left to right) [Y/H], [Zr/H], [Ba/H], [La/H], [Ce/H], [Nd/H], and [Eu/H] reported by \citetalias{Schiappacasse-Ulloa2024}, \citet{Carretta2013,Carretta2011,Carretta2014} and the {\it Gaia}-ESO {\sc dr5.1} for in-common stars. The figure follows the same description as Fig.\ref{fig:abund_carr}.}
        \label{fig:abund_mine}
\end{figure*}

Finally, Fig.\ref{fig:abund_mine} shows the comparison of n-capture elements reported by \citetalias{Schiappacasse-Ulloa2024} and \citet{Carretta2011,Carretta2013,Carretta2014}, when available. The figure reveals that both {\it Gaia}-ESO {\sc DR5.1} and the literature follow the same trend for the most populated species-- however -- with a larger spread with respect to the lighter elements. The larger difference is reported for the abundances of [Y/H], [Ba/H], [Eu/H], where the {\it Gaia}-ESO {\sc dr5.1} found higher abundances with respect to the literature. The difference is likely primarily due to the high sensitivity of Y and Ba abundances to stellar parameters (the clear \teff ~offset and the different \vm ~distribution), with an additional contribution from the different line lists used.

\section{Effect of \vm on Y and Ba in our GCs sample}

\begin{figure*}
        \centering
        \includegraphics[width=1.3\columnwidth]{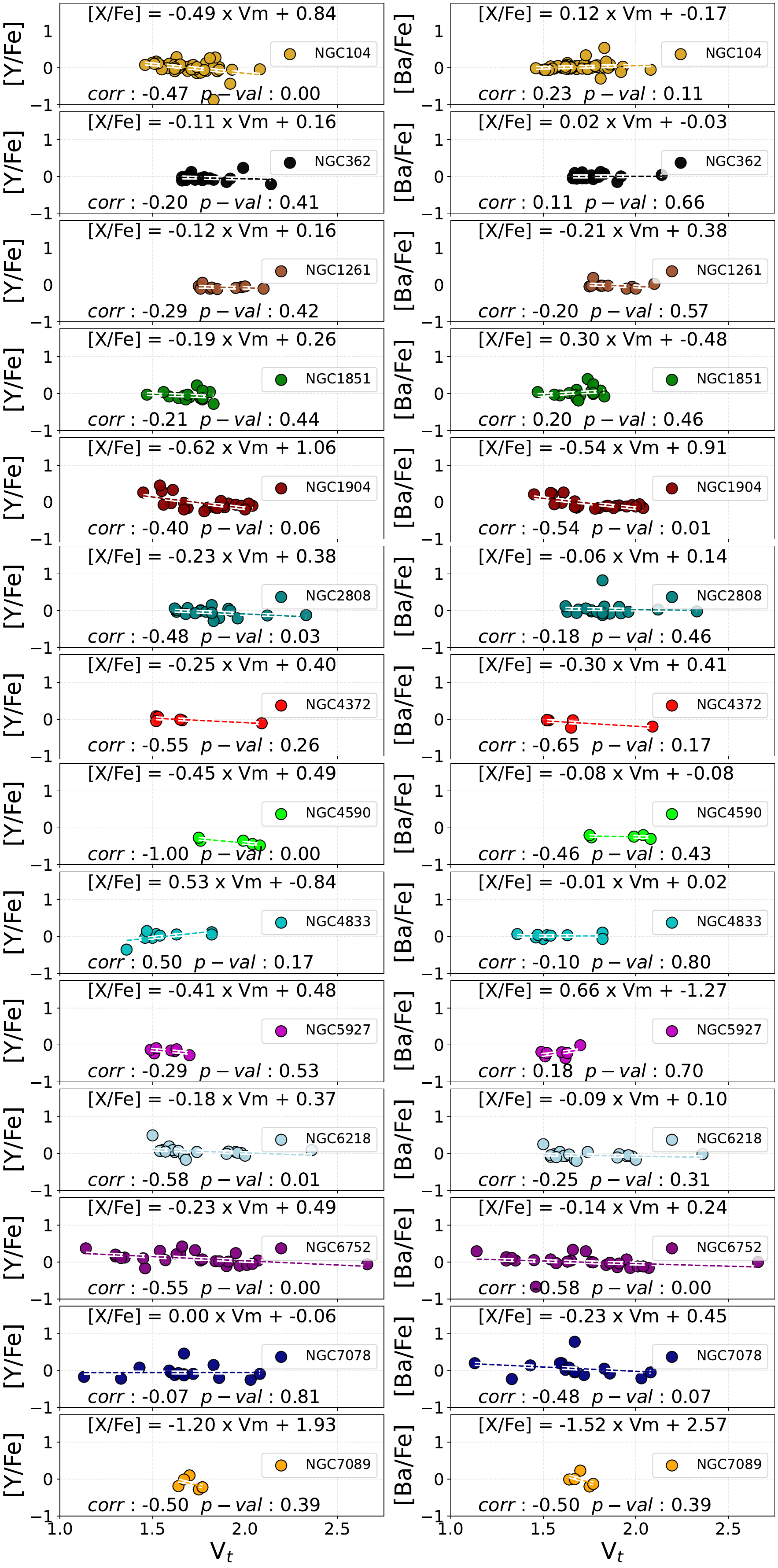}
        \caption{[Y/Fe] and [Ba/Fe] abundances as a function of \vm for the whole sample of GCs. Every panels displays the one-degree fit's equation and the respective Spearman correlation and p-value.}
        \label{fig:YBA_Vm}
    \end{figure*}

As mentioned in \S \ref{Sec:YBA_Vm}, strong lines such as those of Y and Ba are often saturated and highly sensitive to \vm, thus leading to a wrong abundance determination of these species. To illustrate how this trend affects the whole sample of GCs, \ref{fig:YBA_Vm} displays the abundances of [Y/Fe] (panels on the left) and [Ba/Fe] (panels on the right) as a function of \vm for each GC. Additionally, each panel displays the corresponding linear fit equation and Spearman statistics. As shown in the figure, some clusters, such as NGC~1904, exhibit a strong dependence of both yttrium and barium abundances on \vm, with slopes of -0.62 and -0.54, respectively. This leads to internal cluster variations of approximately 0.50 dex in [Y/Fe] and 0.35 dex in [Ba/Fe] within its \vm ~range. In contrast, other clusters exhibit negligible slopes, suggesting minimal or no dependence on \vm. A quick comparison between two GCs—one with a strong Y trend (NGC~1904) and one with a weak Y trend (NGC~7078)—reveals that the cluster with the stronger trend also spans a wider range of \teff ~and $\log g$.

\section{Additional n-capture elements and MP indicators trends}
\label{App:ncap-MP}

\begin{figure}
        \centering
        \includegraphics[width=\columnwidth]{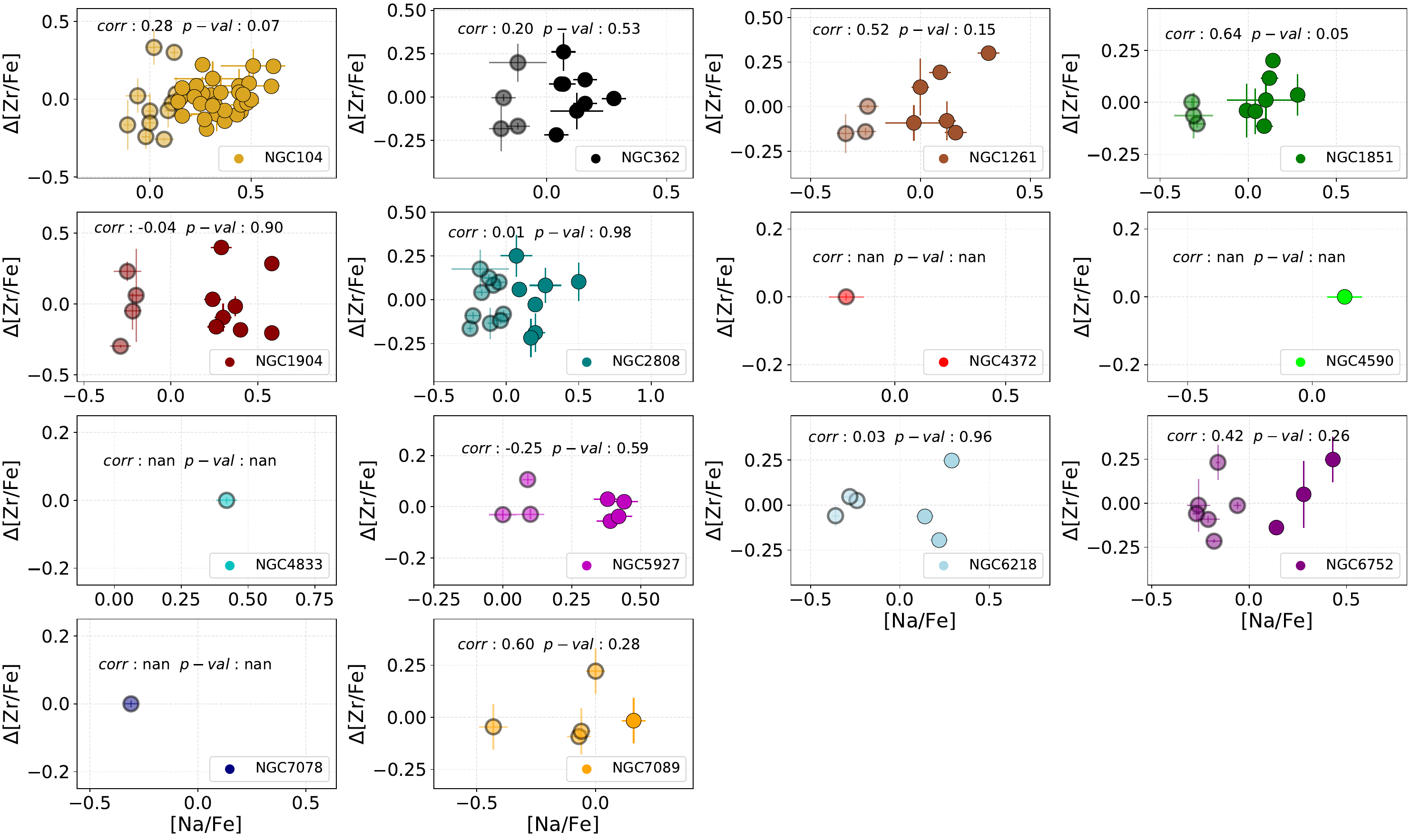}
        \includegraphics[width=\columnwidth]{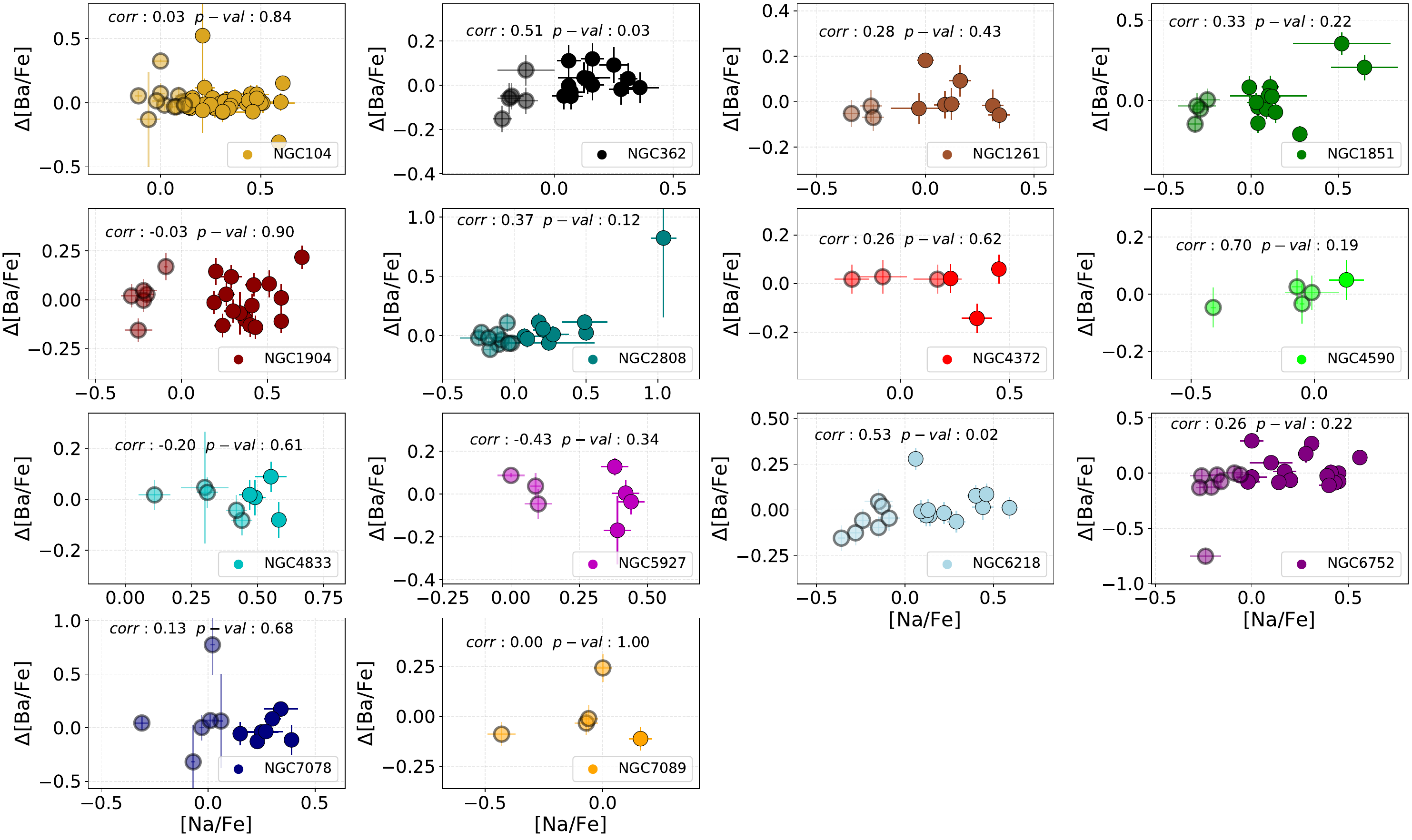}
        \includegraphics[width=\columnwidth]{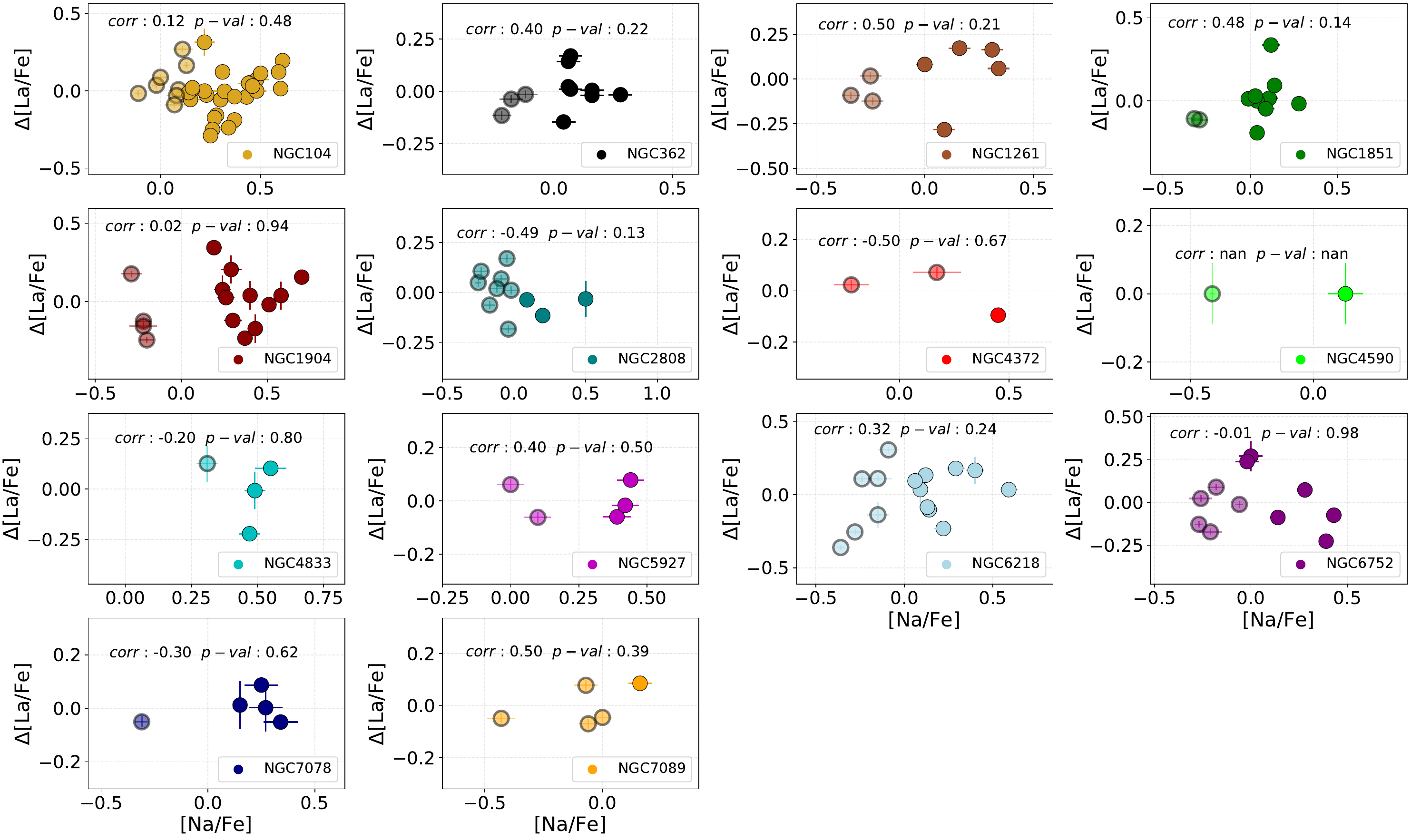}
        \caption{Abundances of $\Delta$[Zr/Fe], $\Delta$[Ba/Fe], and $\Delta$[La/Fe] as a function of [Na/Fe] for GCs in our sample. Every panel shows the correspondent Spearman correlation and p-value.}
        \label{fig:corr_na1}
\end{figure}

\begin{figure}
        \centering
        \includegraphics[width=\columnwidth]{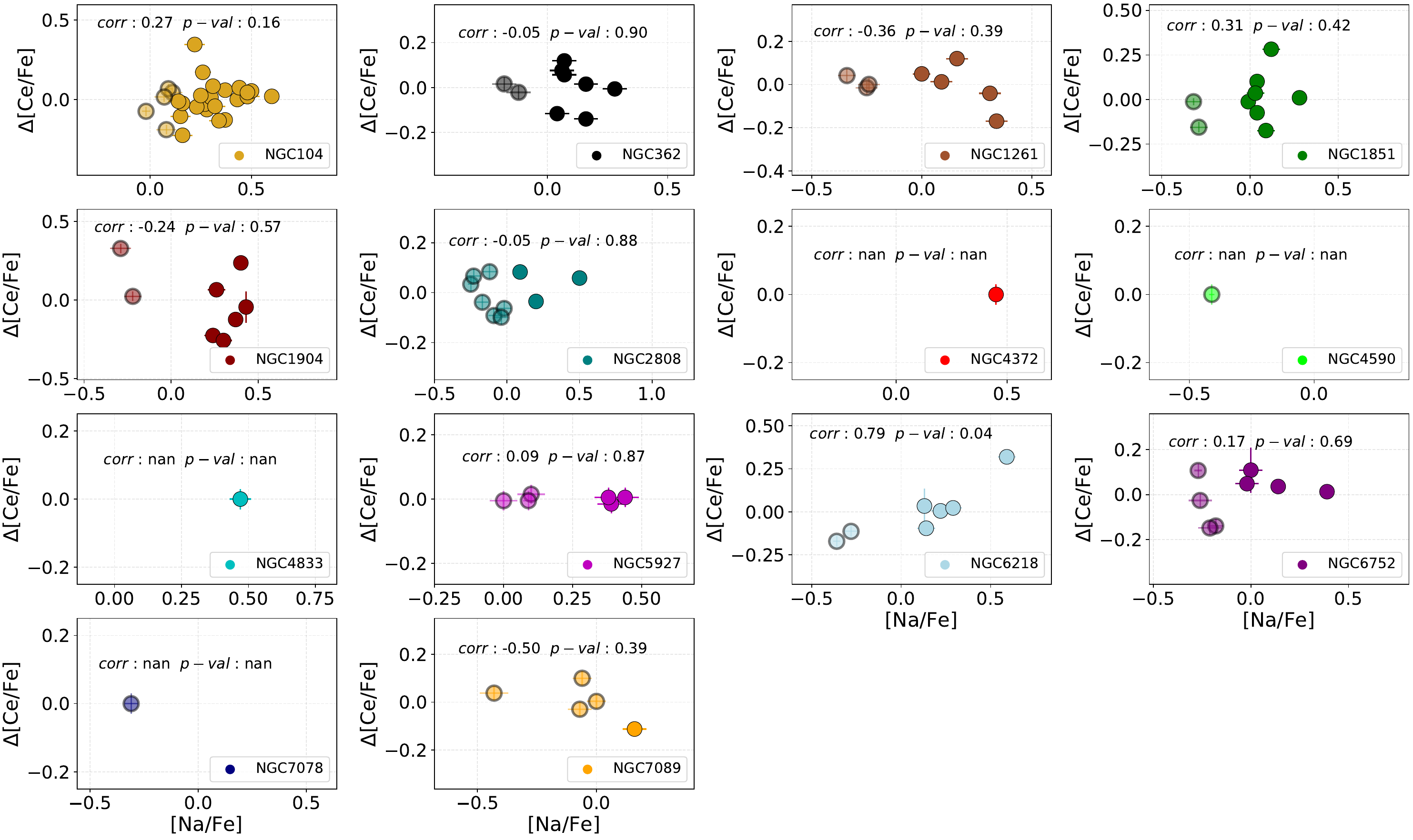}
        \includegraphics[width=\columnwidth]{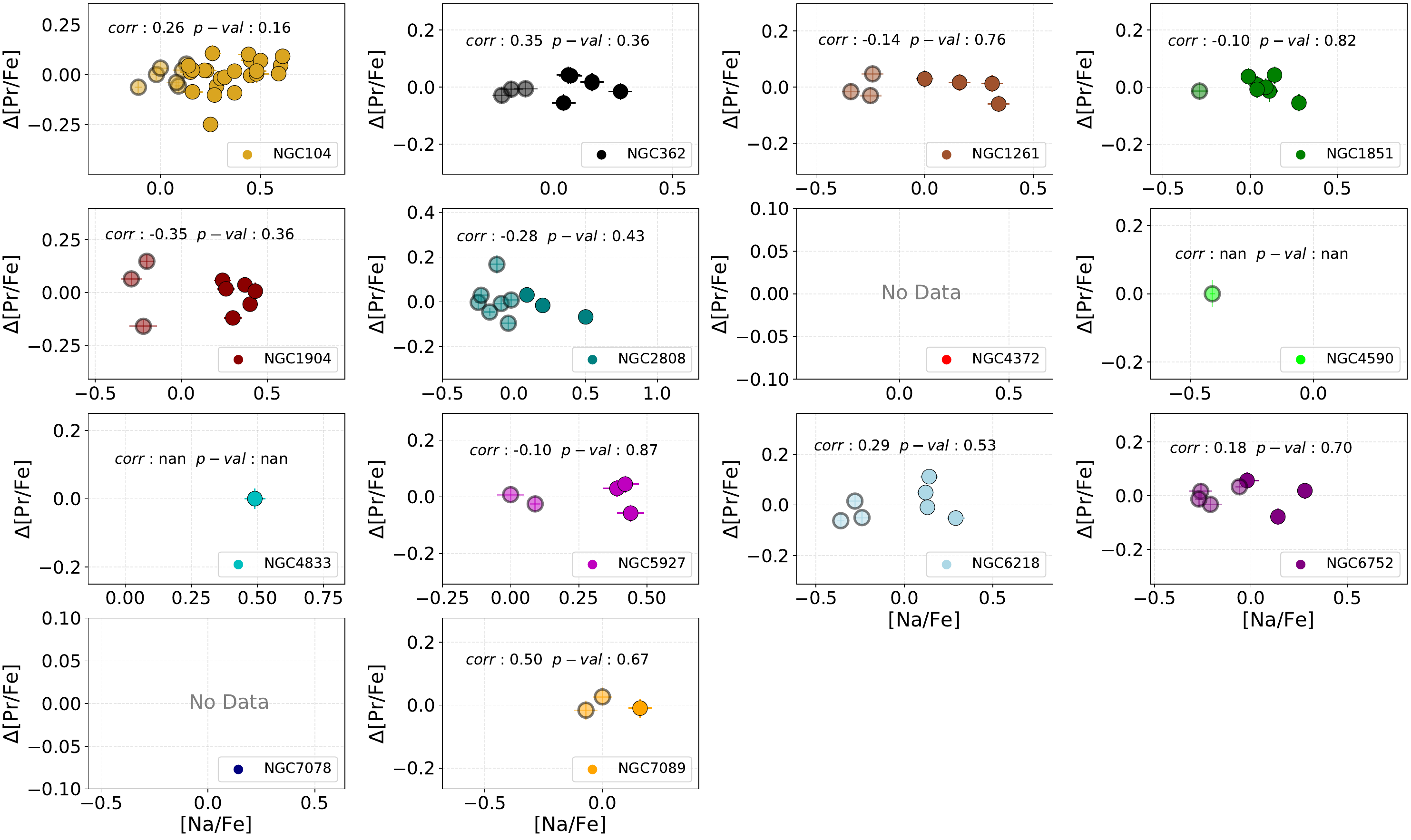}
        \includegraphics[width=\columnwidth]{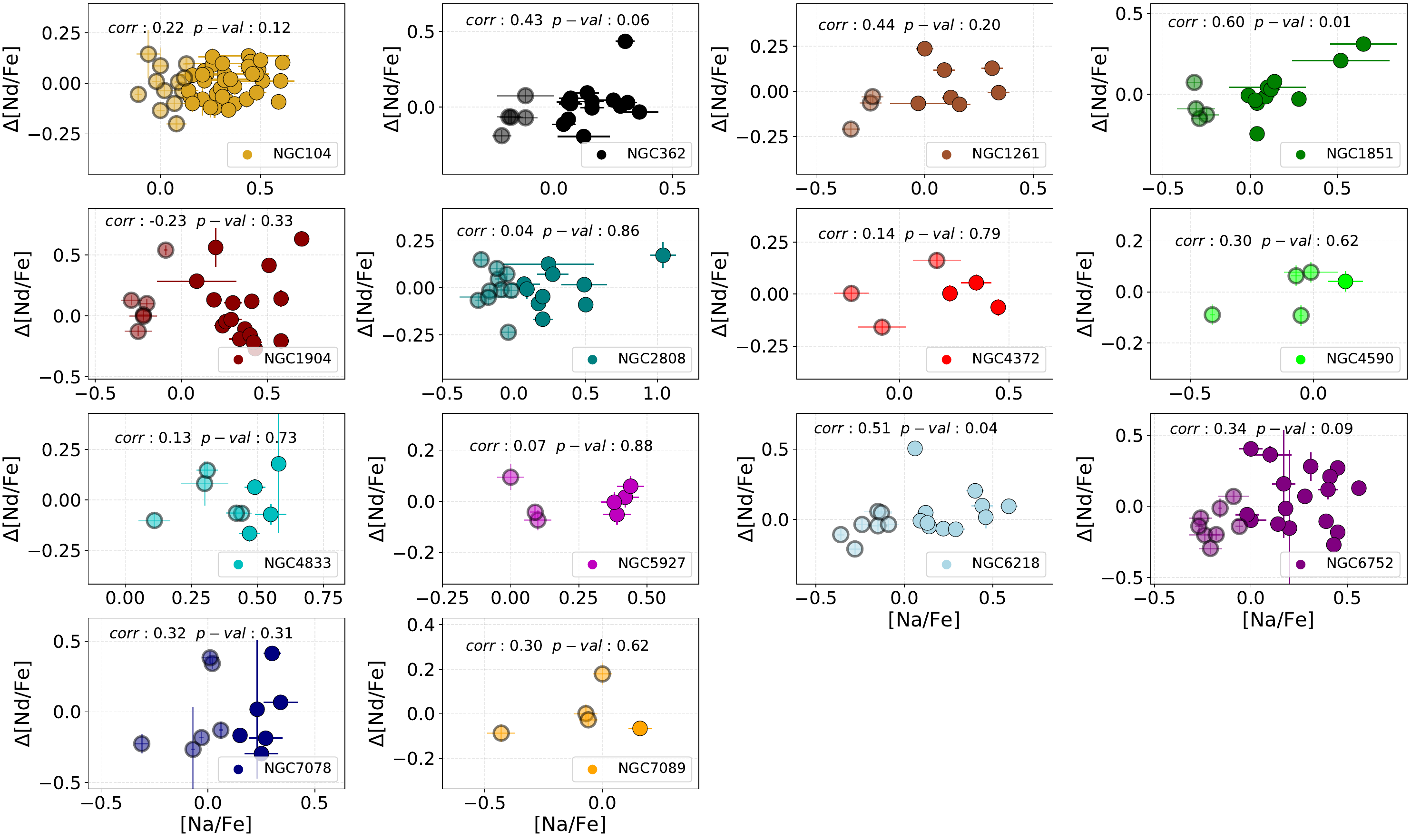}
        \includegraphics[width=\columnwidth]{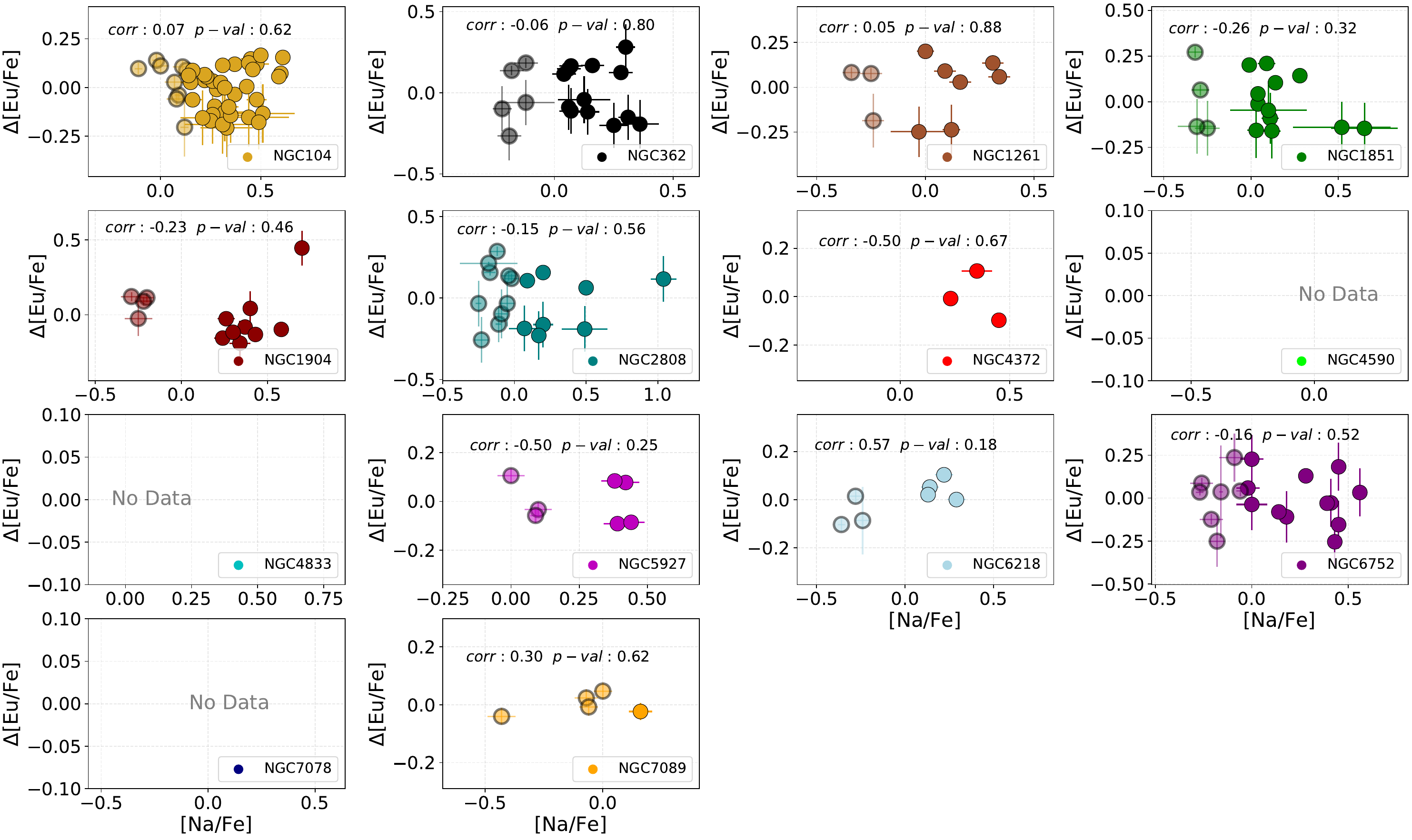}
        \caption{Abundances of $\Delta$[Ce/Fe], $\Delta$[Pr/Fe], $\Delta$[Nd/Fe], and $\Delta$[Eu/Fe] as a function of [Na/Fe] for GCs in our sample. Every panel shows the correspondent Spearman correlation and p-value.}
        \label{fig:corr_na2}
\end{figure}

\begin{figure}
        \centering
        \includegraphics[width=\columnwidth]{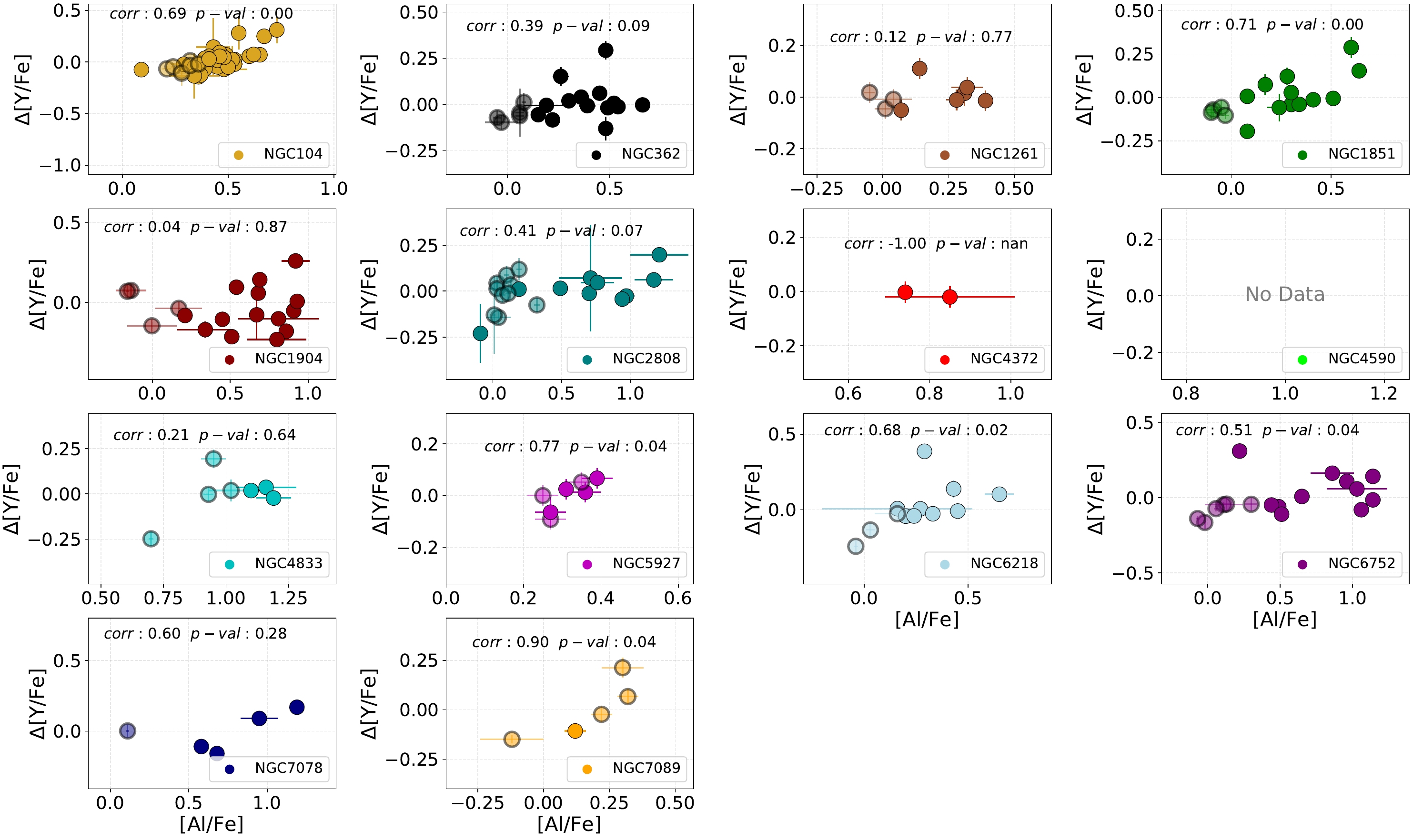}
        \includegraphics[width=\columnwidth]{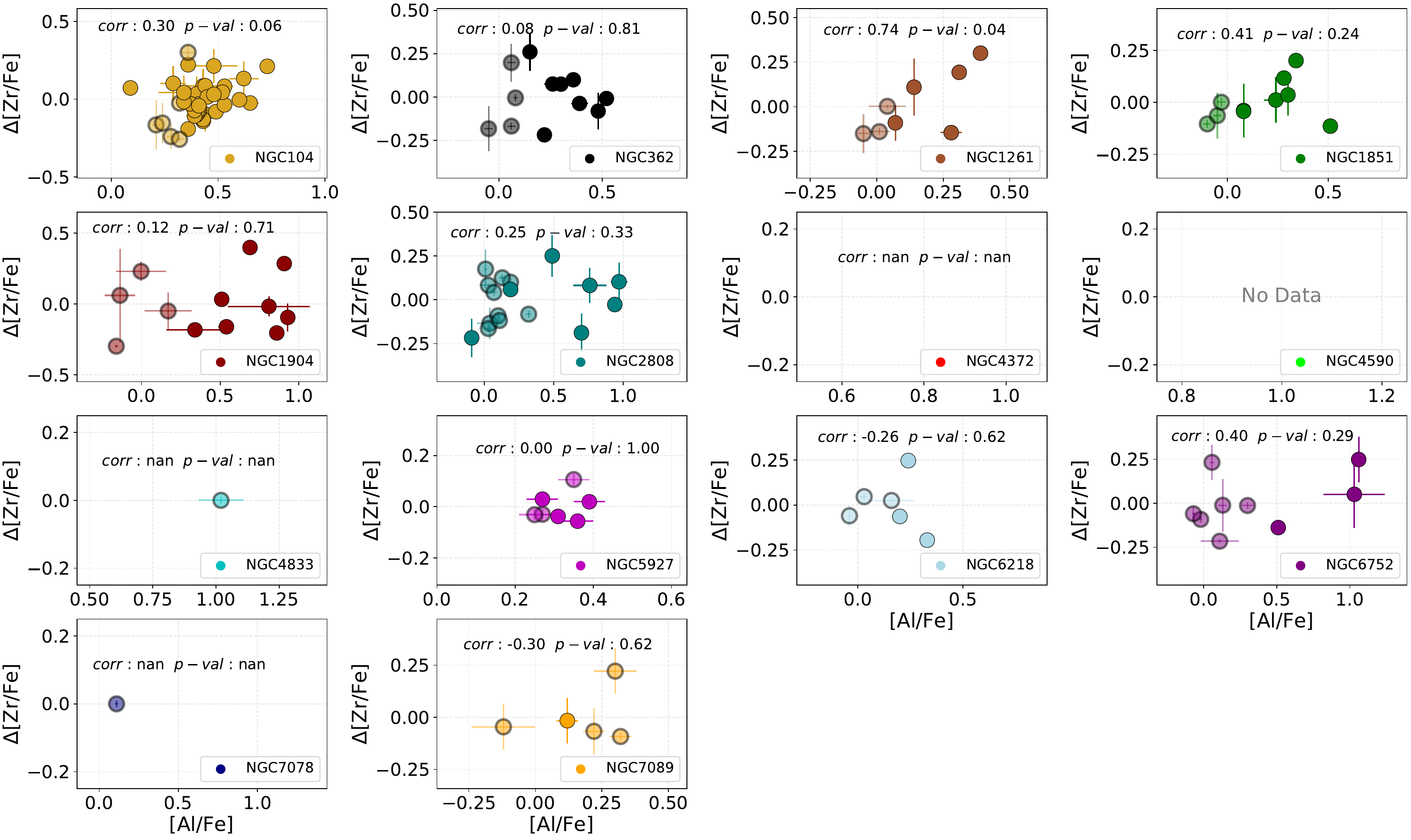}
        \includegraphics[width=\columnwidth]{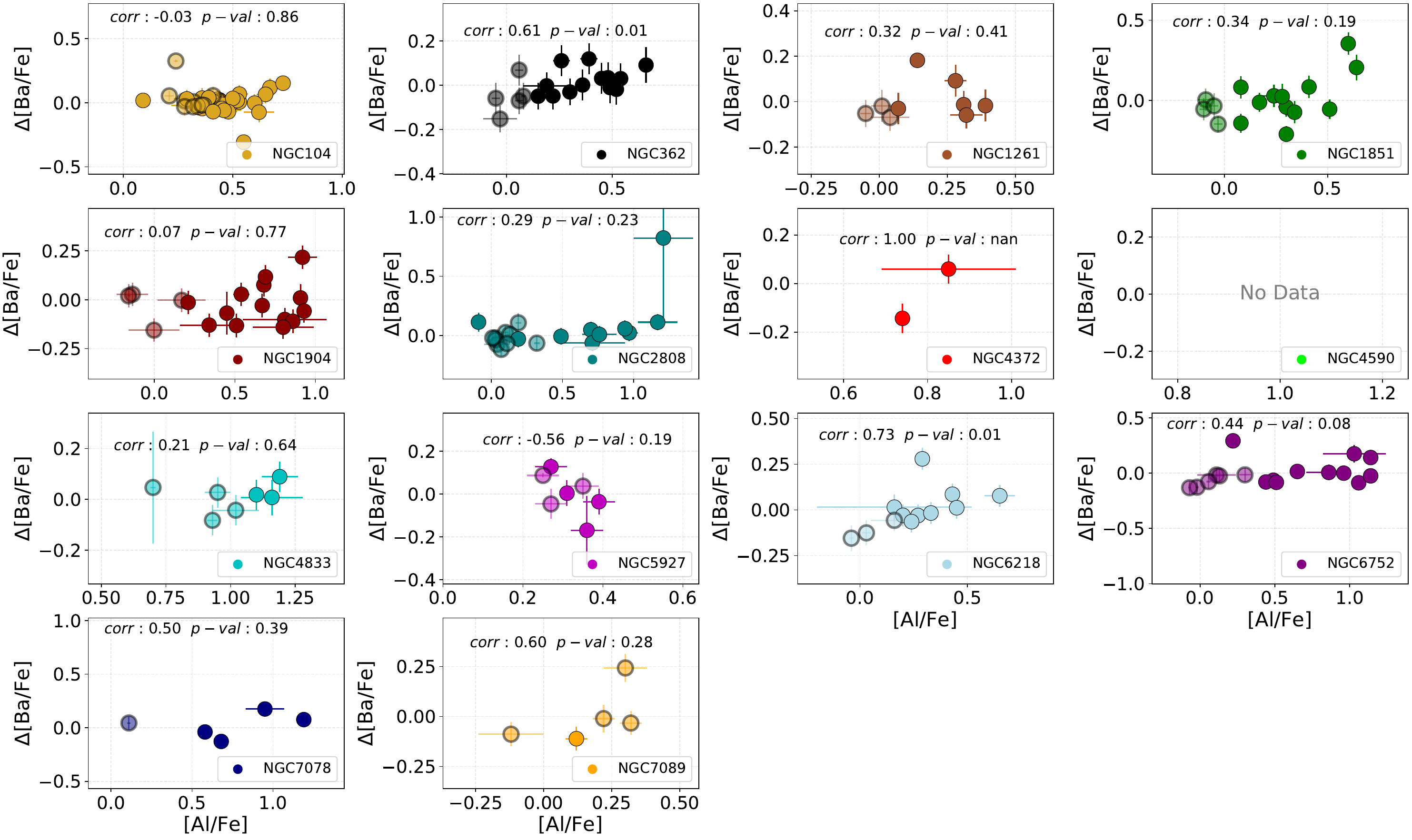}
        \includegraphics[width=\columnwidth]{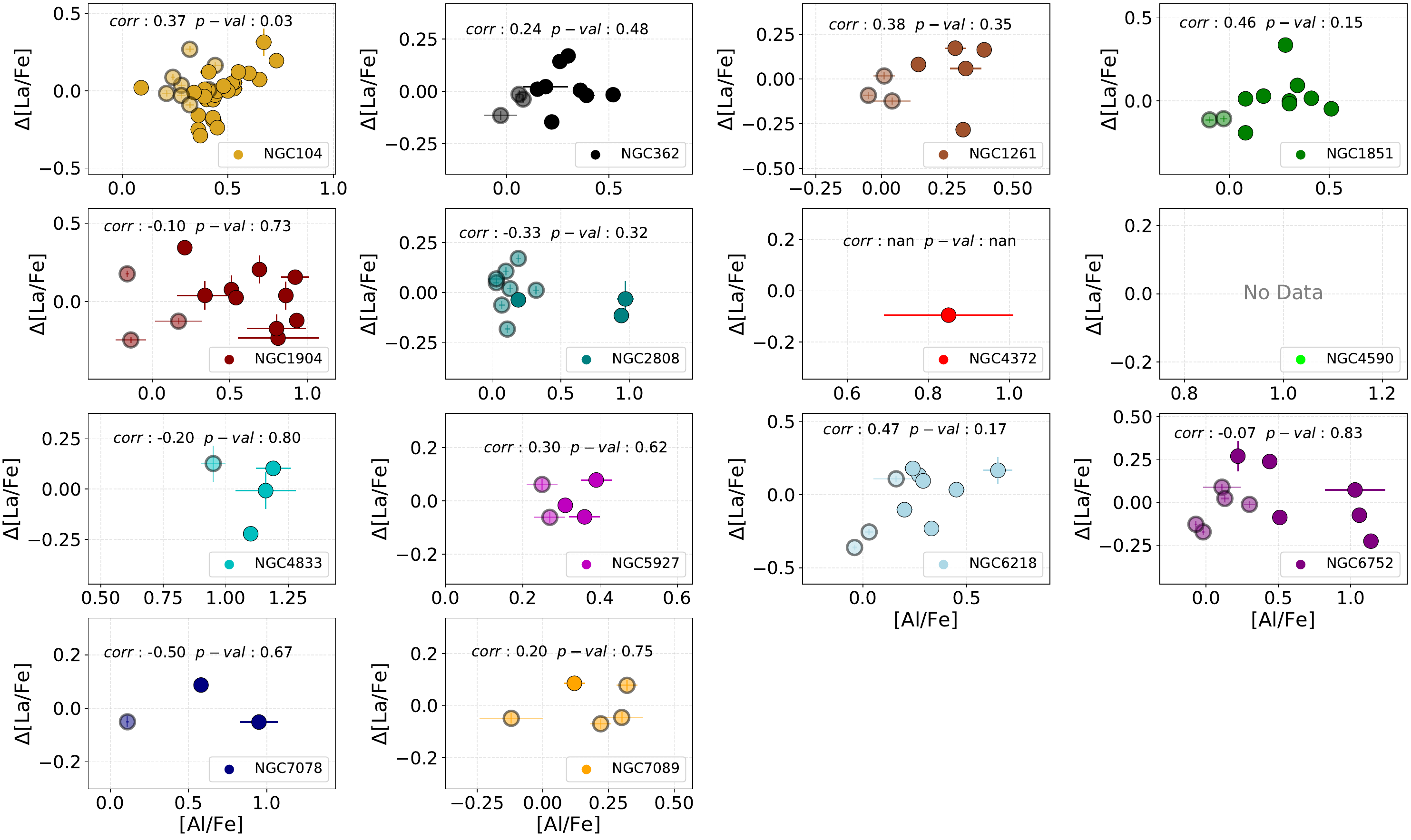}
        \caption{Abundances of $\Delta$[Y/Fe], $\Delta$[Zr/Fe], $\Delta$[Ba/Fe], and $\Delta$[La/Fe] as a function of [Al/Fe] for GCs in our sample. Every panel shows the correspondent Spearman correlation and p-value.}
        \label{fig:corr_al1}
\end{figure}

\begin{figure}
        \centering
        \includegraphics[width=\columnwidth]{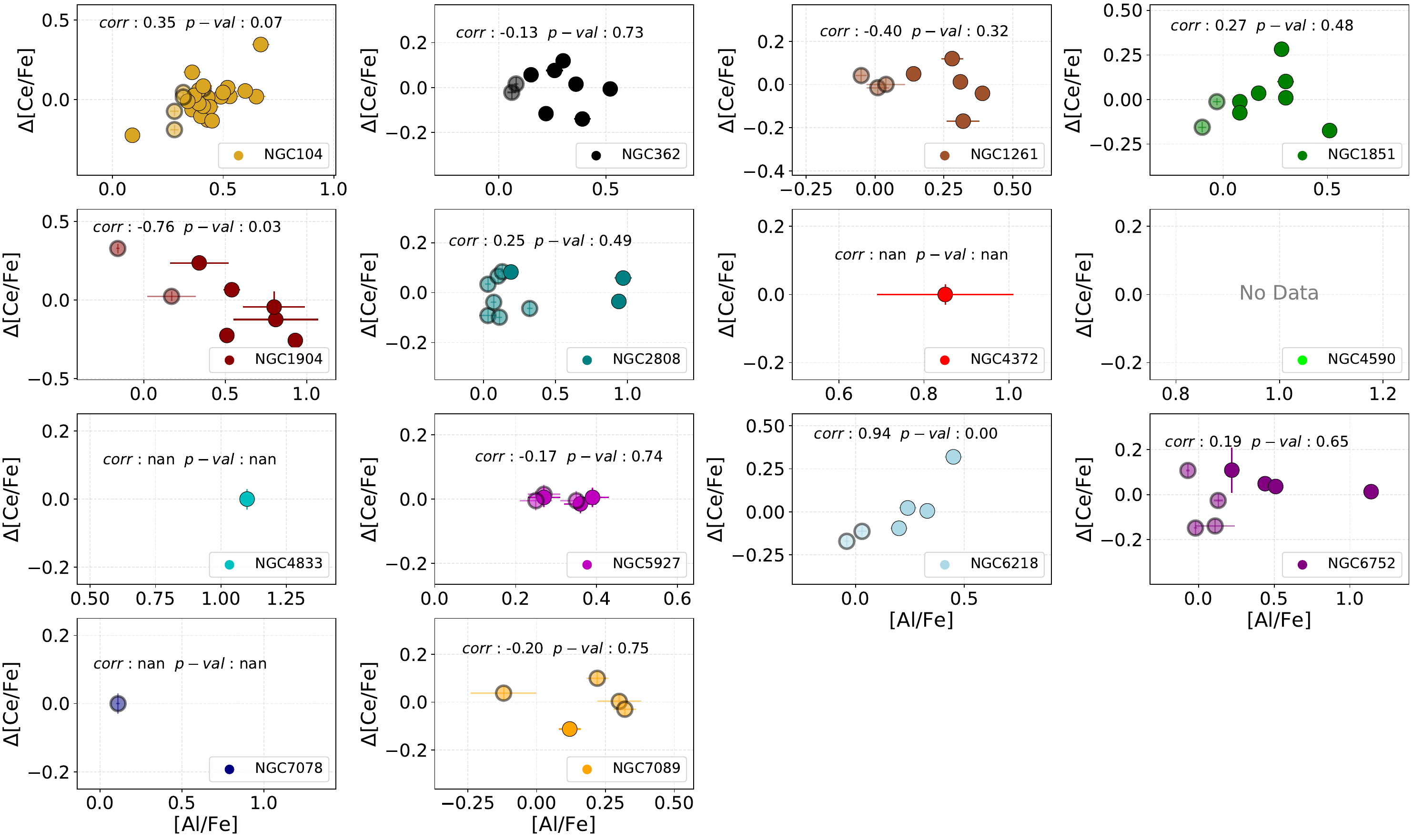}
        \includegraphics[width=\columnwidth]{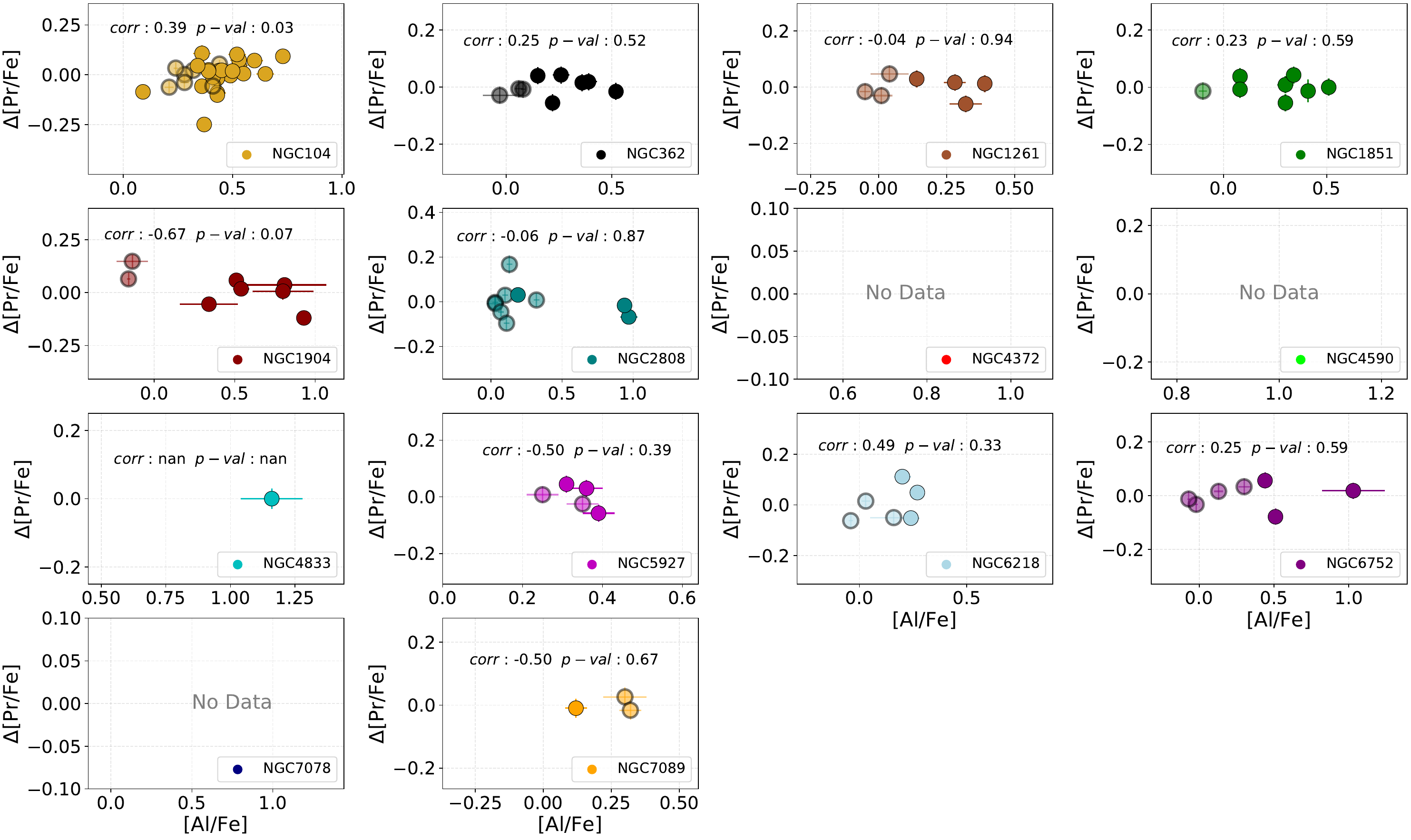}
        \includegraphics[width=\columnwidth]{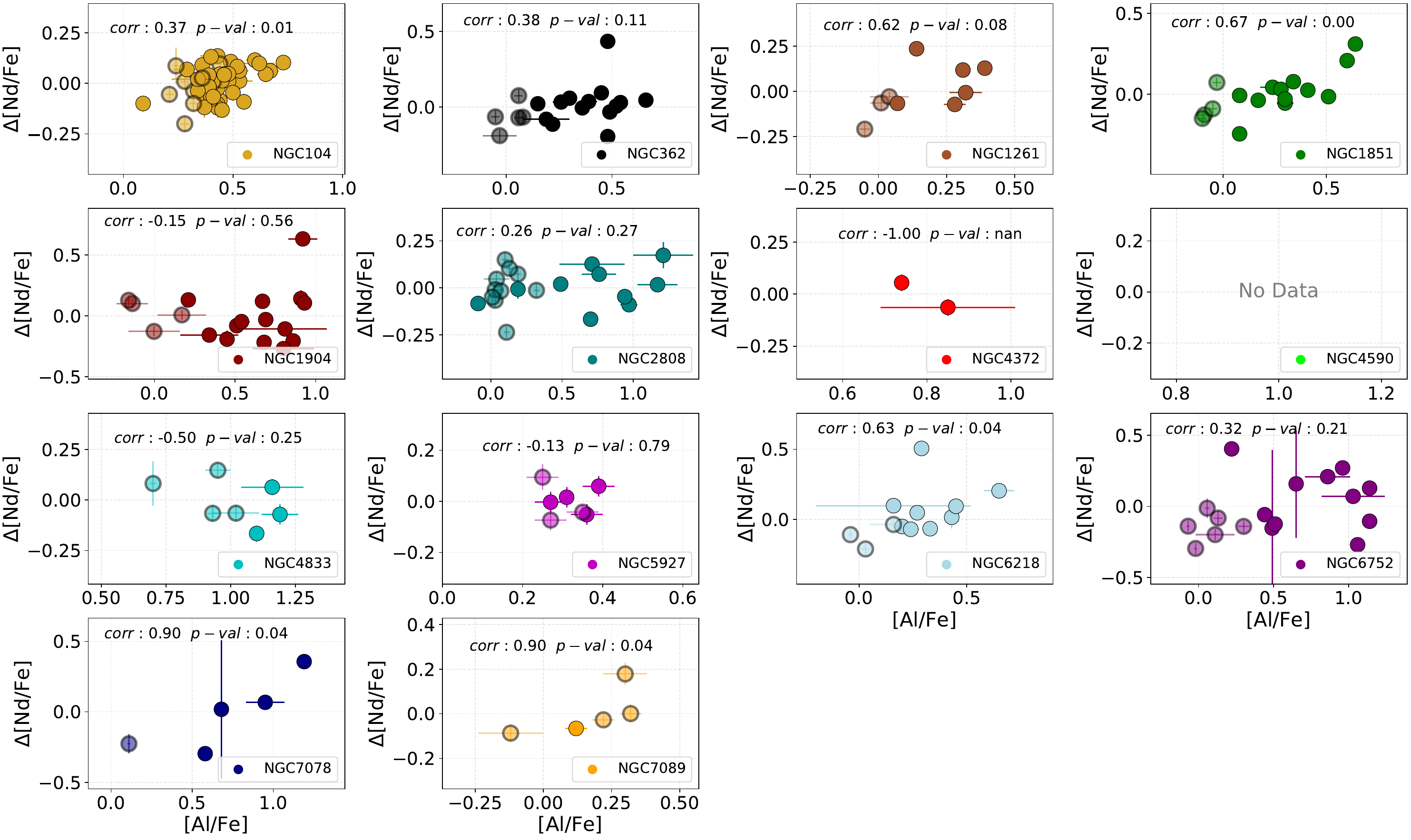}
        \includegraphics[width=\columnwidth]{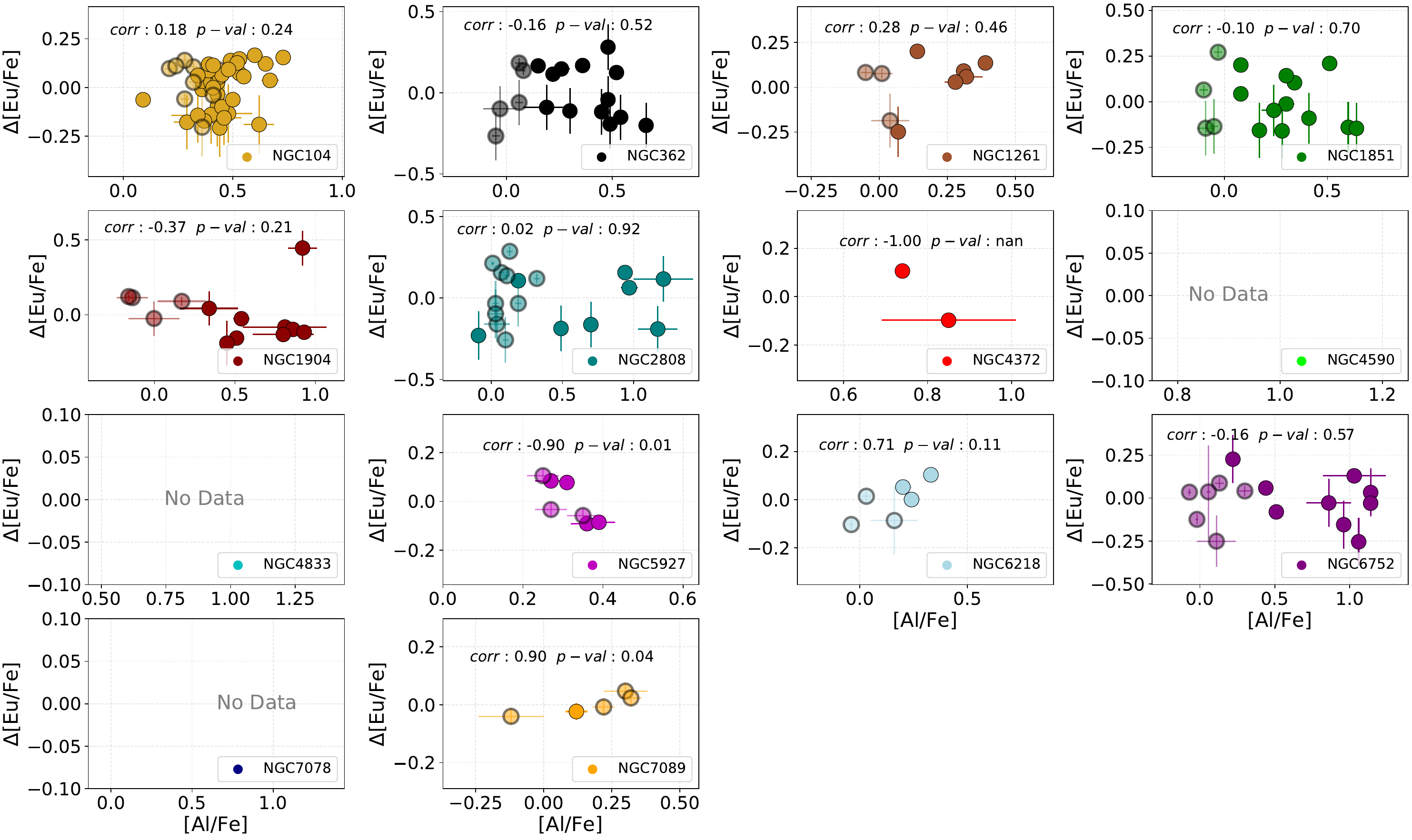}
        \caption{Abundances of $\Delta$[Ce/Fe], $\Delta$[Pr/Fe], $\Delta$[Nd/Fe], and $\Delta$[Eu/Fe] as a function of [Al/Fe] for GCs in our sample. Every panel shows the correspondent Spearman correlation and p-value. }
        \label{fig:corr_al2}
\end{figure}

As discussed in \S\ref{Sec:n-cap_vs_NAL}, we examined possible correlations between the n-capture elements analysed in this work and the commonly used MP indicators Na and Al. Figures \ref{fig:corr_na1} and \ref{fig:corr_na2} show the abundances of various n-capture elements as a function of Na, while Figures \ref{fig:corr_al1} and \ref{fig:corr_al2} present the corresponding plots for Al. A detailed interpretation of these trends is provided in \S\ref{Sec:n-cap_vs_NAL}.

\label{App:vm}

\end{appendix}

\end{document}